\documentclass[fleqn,usenatbib]{mnras}

\usepackage{newtxmath}

\usepackage[T1]{fontenc}
\usepackage{ae,aecompl}

\usepackage{graphicx}	
\usepackage{amsmath}	
\usepackage{amssymb}	

\usepackage[dvipsnames]{xcolor}
\hypersetup{
  colorlinks   = true, 
  urlcolor     = RoyalBlue, 
  linkcolor    = RoyalBlue, 
  citecolor    = MidnightBlue 
}
\usepackage{natbib}
\usepackage{color}
\usepackage{rotating}
\usepackage{url}
\usepackage{ifthen}

\usepackage{bm}
\usepackage{bbold}
\usepackage{aas_macros}
\usepackage{verbatim}
\usepackage{algorithm}
\usepackage{algpseudocode}
\usepackage{enumitem}
\usepackage{subfig}

\usepackage{tablefootnote}
\usepackage[flushleft]{threeparttable}

\DeclareMathAlphabet{\mathpzc}{OT1}{pzc}{m}{it}

\newcommand{\mvec}[1]{\bm{#1}}

\newcommand{\myvec}[1]{\mvec{#1}}

\newcommand\numberthis{\addtocounter{equation}{1}\tag{\theequation}}

\SetSymbolFont{symbols}{bold}{OMS}{cmsy}{b}{n}
\DeclareSymbolFont{bmisymbols}{OML}{cmm}{b}{it}

\title[Dynamical mass inference with neural flows]{Dynamical mass inference of galaxy clusters with neural flows}

\author[D. Kodi Ramanah et al.]{Doogesh Kodi Ramanah,$^{1}$\thanks{ramanah@nbi.ku.dk} Rados\l{}aw Wojtak,$^{1}$ Zoe Ansari,$^{1}$ Christa Gall,$^{1}$\newauthor Jens Hjorth$^{1}$\\
$^{1}$ DARK, Niels Bohr Institute, University of Copenhagen, Jagtvej 128, 2200 Copenhagen, Denmark
}

\date{Accepted XXX. Received YYY; in original form ZZZ}

\pubyear{2020}

\begin{document}
\label{firstpage}
\pagerange{\pageref{firstpage}--\pageref{lastpage}}
\maketitle

\begin{abstract}
We present an algorithm for inferring the dynamical mass of galaxy clusters directly from their respective phase-space distributions, i.e. the observed line-of-sight velocities and projected distances of galaxies from the cluster centre. Our method employs normalizing flows, a deep neural network capable of learning arbitrary high-dimensional probability distributions, and inherently accounts, to an adequate extent, for the presence of interloper galaxies which are not bounded to a given cluster, the primary contaminant of dynamical mass measurements. We validate and showcase the performance of our neural flow approach to robustly infer the dynamical mass of clusters from a realistic mock cluster catalogue. A key aspect of our novel algorithm is that it yields the probability density function of the mass of a particular cluster, thereby providing a principled way of quantifying uncertainties, in contrast to conventional machine learning approaches. The neural network mass predictions, when applied to a contaminated catalogue with interlopers, have a mean overall logarithmic residual scatter of 0.028~dex, with a log-normal scatter of 0.126~dex, which goes down to 0.089~dex for clusters in the intermediate to high mass range. This is an improvement by nearly a factor of four relative to the classical cluster mass scaling relation with the velocity dispersion, and outperforms recently proposed machine learning approaches. We also apply our neural flow mass estimator to a compilation of galaxy observations of some well-studied clusters with robust dynamical mass estimates, further substantiating the efficacy of our algorithm.
\end{abstract}




\section{Introduction}
\label{intro}

The abundance of galaxy clusters is one of the most fundamental predictions of cosmological model and structure formation \citep{Kravtsov2012}. Counts of galaxy clusters as a function of dynamical mass are traditionally used to constrain the properties of dark matter \citep{Rozo2010,Bocquet2019} and dark energy \citep{Vikhlinin2009,Mantz2015}, and test Einstein's theory of general relativity via the measurements of the linear growth \citep{Rapetti2010}. This is where the accuracy and precision of cluster mass estimates directly impact the robustness of cosmological inference, providing strong motivation for developing new techniques capable of handling constantly growing observational data.

\medskip
Upcoming large-scale imaging and spectroscopic surveys, such as Dark Energy Spectroscopic Instrument \citep[DESI;][]{DESI2016}, the Vera C. Rubin Observatory \citep{lsst2008summary} and Euclid \citep{euclid2016missiondesign}, will open up unprecedented opportunities for cluster cosmology with the potential of measuring the dark energy equation of state parameter to percent precision \citep{Sartoris2016}. The surveys will revive and challenge traditional methods of inferring dynamical cluster masses from relatively simple observables comprising the projected positions and velocities of cluster galaxies. The vast amount of future observations as well as the ongoing development of mock data from cosmological simulations will enable more effective dynamical mass inference based on neural network algorithms trained on mock observations accounting for all relevant selection effects.

\medskip
Masses of $N$-body systems have been traditionally measured from kinematic data (projected positions and velocities) using methods based on the virial theorem \citep[e.g.][]{heisler1985estimating}. Keeping the core assumption of dynamical equilibrium and spherical symmetry, more complex approaches were successively developed. More general mass estimators derived from the Jeans equation allowed for incorporating broader assumptions about the underlying density profiles and the orbital anisotropy \citep{an2011modified}. The increasing amount of observations enabled fitting complete solutions of the Jeans equation to the observed radial profiles of the projected velocity dispersion. This approach allowed constraining more detailed features of the systems such as density profiles or the orbital anisotropy \citep{Biviano2004}. The Jeans modelling was finally extended by including higher velocity moments \citep{lokas2003coma} or by complete models describing the distribution of galaxies in the projected phase space \citep{Wojtak2009, mamon2013mamposst}. This effectively exhausted theoretical possibilities permitted under the assumption of dynamical equilibrium and spherical symmetry.

\medskip
Dynamical equilibrium and spherical models are highly idealized assumptions. They are not fully applicable to galaxy clusters which are not in perfect equilibrium \citep{natarajan1997distribution}, surrounded by infalling structures \citep{Gunn1972} and quite often exhibiting signatures of recent mergers. A way to circumvent this problem is to use non-equilibrium models linking the observed infall velocities to dynamical mass, e.g. the caustic method \citep{diaferio1997infall} or the Jeans equation with non-vanishing streaming motion \citep{Falco2014}. Other approaches are based on exploiting phenomenological scaling relation between cluster dynamical mass and some easily measurable properties such as cluster richness \citep{Rykoff2012}. Further complications in dynamical mass estimations can arise from the fact that galaxy clusters are not spherical, neither in position space \citep{deTheije1995} nor in velocity space \citep{wojtak2013phase}. This can likely degrade precision of the mass estimators based on kinematic data \citep{svensmark2015effect}. Additional systematic errors are expected to be caused by substructures \citep{old2018clusterIII} and the presence of interlopers \citep{wojtak2018clusterIV}. The latter effect appears to be the primary factor degrading the quality of cluster mass estimates. According to several tests performed on mock observations of galaxy clusters, dynamical masses or rich clusters with $\sim10^{2}$ member galaxies can be currently measured to a precision no better than $0.15$~dex, quite often with a mass-dependent accuracy \citep{old2015clusterII, armitage2019eagle}, while the minimum attainable uncertainty for ideal measurements free of systematic effects related to imperfect interloper removal can be as low as $0.05$~dex \citep{wojtak2018clusterIV}.

\medskip
Machine learning (ML) approaches are becoming increasingly popular for applications involving the estimation of dynamical cluster masses. The initial implementations relied on non-parametric methods, where an ML model is optimized using a large training data set typically consisting of simulated mock observations. Once trained, the ML model may then be utilized to estimate cluster masses from unlabelled data sets. For instance, \citet{ntampaka2015machine, ntampaka2016dynamical} used support distribution machines \citep{sutherland2012SDM} to directly infer the cluster mass from line-of-sight velocities and positions of galaxies. \citet{armitage2019application} recently implemented a series of simple ML regression models, such as linear, ridge and kernel ridge regression, on a set of features extracted manually from cluster observations. This was followed by the work of \citet{calderon2019prediction}, who implemented three more complex ML algorithms, namely XGBoost, random forests and neural network. The above ML approaches all led to similar significant improvements relative to the standard virial ($M-\sigma$) scaling relation, effectively reducing the prediction scatter by a factor of two, thereby highlighting the potential of ML-based approaches as promising alternatives to classical methods of cluster mass estimation.

\medskip
Convolutional neural networks (CNNs) were recently designed by \cite{ho2019robust} to further improve upon the previous ML approaches. Since CNNs are especially effective for image recognition purposes, they made use of a kernel density estimator to generate 2D phase-space maps of individual cluster dynamics, which are fed to the CNN as input images. The training rationale is a standard regression over logarithmic cluster mass with a realistic mock simulation providing the training and test sets. This CNN approach yields an overall improvement by a factor of three over the classical $M-\sigma$ estimators, with the network prediction scatter reducing to $\sim 0.132$ dex for the more massive clusters. \cite{ho2019robust} also provides an excellent in-depth review of the above different classes of ML algorithms applied to cluster mass measurements.

\medskip
In this work, we present a novel dynamical mass inference algorithm, inspired by the recently developed framework of neural flows \citep{rezende2015normalizing, germain2015made, papamakarios2017MAF}, with our method being complementary to the ML approaches outlined above. Additionally, however, our neural flow model yields the conditional probability distribution of the dynamical mass of individual clusters, given their respective phase-space kinematics, thereby providing a way to quantify uncertainties, rather than predicting only single point estimates. Our training, validation and test data sets are drawn from the simulated cluster catalogue from \citet{ho2019robust}, which emulates the physical artefacts encountered in practice, described in Section~\ref{mock_catalogues}.

\medskip
The remainder of this paper is structured as follows. In Section~\ref{mass_estimation_problem}, we briefly describe the general problem of mass estimation of galaxy clusters and outline the generation of the mock galaxy cluster catalogues employed in the training and validation of our neural network. We review the conceptual underpinnings of the neural density estimators in Section~\ref{neural_density_estimators}, with particular emphasis on normalizing flows, followed by a description of our network architecture and training procedure in Section~\ref{neural_flows}. We subsequently validate and showcase the performance of our neural network in Section~\ref{validation_performance}, and follow-up by illustrating a few applications on real galaxy cluster data sets in Section~\ref{applications}. In Section~\ref{model_interpretation}, we derive saliency maps in an attempt to introspect the model performance. Finally, in Section~\ref{conclusions}, we summarize the salient aspects of our work and outline potential extensions to further refine our neural flow approach.

\section{Estimating dynamical mass of galaxy clusters}
\label{mass_estimation_problem}

In this section, we provide a brief overview of the general problem of cluster mass estimation, including the virial scaling relation, and outline the generation of the mock cluster catalogues used in subsequent sections.

\subsection{Classical $M - \sigma$ relation}
\label{classical_m_sigma}

The virial scaling relation provides a simplified means to estimate the cluster mass $M$ via only a single summary statistic, the so-called galaxy velocity dispersion $\sigma_{\mathrm{v}}$, and is, hence, referred to as the $M-\sigma$ relation. The classical form of this relation may be derived from the equivalence of kinetic and potential energy, as encoded in the virial theorem, under the assumption of static and spherically symmetric clusters. The $M-\sigma$ relation is typically expressed as
\begin{equation}
    \sigma_{\mathrm{v}} = \sigma_{\mathrm{v}, 15} \left[ \frac{h(z) M_{\mathrm{200c}}}{10^{15} M_{\odot}} \right]^{\alpha} ,
    \label{eq:classical_M_sigma_relation}
\end{equation}
where $M_{\mathrm{200c}}$ is the cluster mass definition adopted in this work, corresponding to the mass enclosed in a spherical region of density $200 \rho_{\mathrm{c}}$, i.e. 200 times the critical density of the Universe, with $M_{\odot}$ denoting solar mass units, $h(z)$ is the dimensionless Hubble rate, and the two scaling parameters: $\sigma_{\mathrm{v}, 15}$ characterizes the velocity dispersion of a cluster with mass $M_{\mathrm{200c}} = 10^{15}h^{-1} M_{\odot}$ and $\alpha$ is the power law scaling exponent globally describing the spatial distribution of mass in a particular cluster. These two scaling factors are generally set to their best-fit values obtained via simulations \citep{evrard2008virial}. The velocity dispersion, as a summary statistic, may be estimated as the standard deviation of galaxy velocities projected along a single line of sight, but this does not capture and exploit all the information from the dynamical phase-space distribution.

\medskip
Galaxy clusters are not perfectly homologous systems. The presence of various effects breaking the assumption of homology gives rise to substantial scatter around the $M - \sigma$ scaling relation. Some examples include physical features, such as dynamical substructure \citep{old2018clusterIII}, cluster triaxiality \citep{svensmark2015effect}, halo environment \citep{white2010cluster} and cluster mergers \citep{ribeiro2011nongaussian}. Moreover, this prediction scatter is exacerbated by selection effects, such as incomplete cluster observations or presence of interlopers \citep{wojtak2018clusterIV}. To mitigate these selection effects when applying the classical $M - \sigma$ relation, complex and sophisticated membership modelling and interloper removal techniques are required \citep[e.g.][]{wojtak2007interloper, mamon2013mamposst, farahi2016galaxy, abdullah2018galweight}.

\subsection{Dynamical phase-space distribution}
\label{dynamical_phase_space_distribution}

\begin{figure}
	\centering
		{\includegraphics[width=0.8\hsize,clip=true]{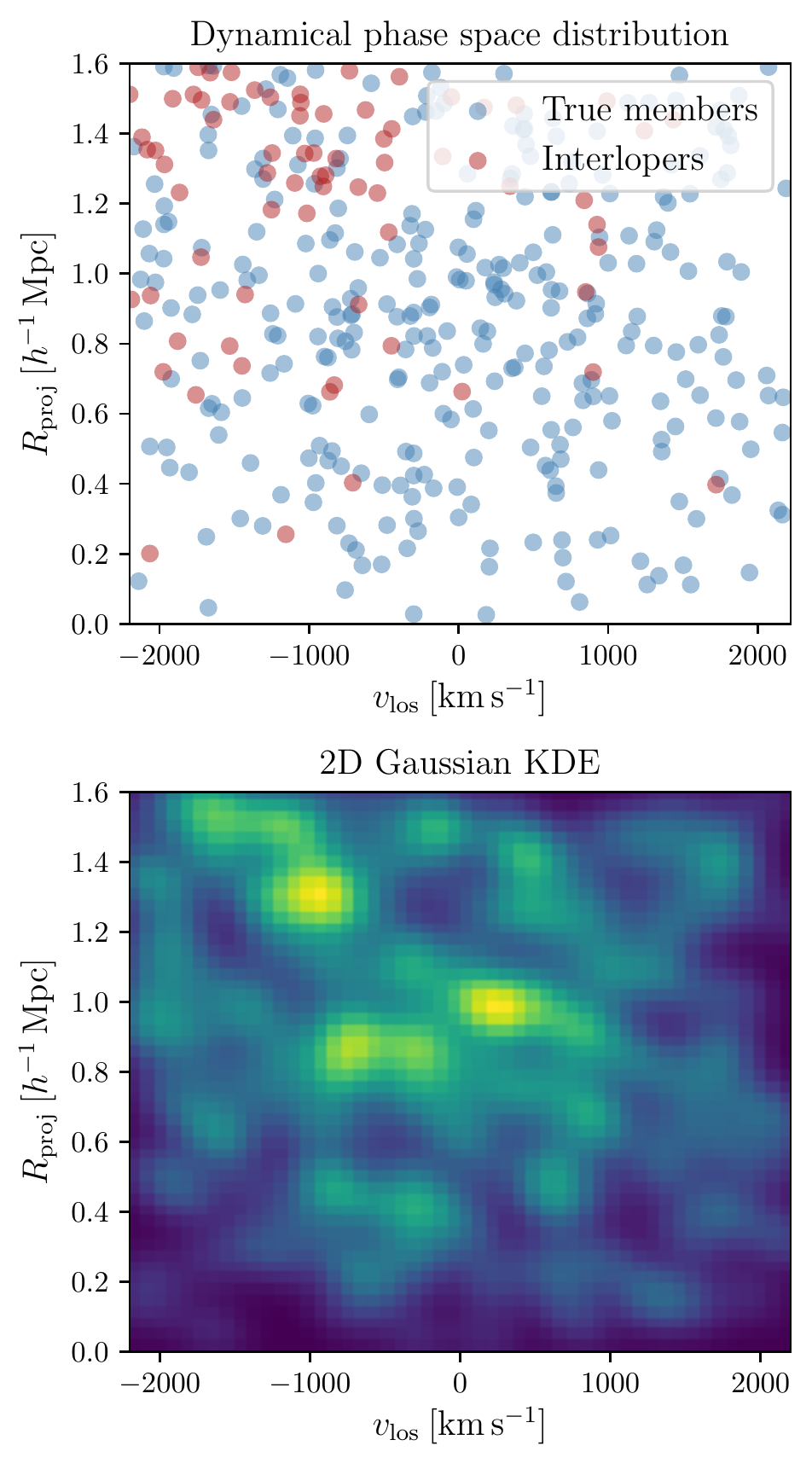}}
	\caption{A sample of a galaxy cluster containing both true members and interloper galaxies. {\it Top panel:} The joint distribution of projected galaxy distances from cluster centre, $R_{\mathrm{proj}}$, and line-of-sight velocities, $v_{\mathrm{los}}$, in dynamical phase space. The true cluster members and interlopers are indicated by blue and red dots, respectively. {\it Bottom panel:} The corresponding 2D Gaussian KDE representation, which serves as inputs to our neural network (cf. Section~\ref{training_methodology}).}
	\label{fig:phase_space_distribution}
\end{figure}

In terms of observations, the galaxy cluster data typically comprise the essential features of the cluster kinematics. For each cluster, the observables are the positions and velocities of its member galaxies to the cluster centre, computed by considering a line-of-sight axis. The positions of all the galaxies around the cluster centre are projected onto the plane $(x, y)$ of the sky and are denoted by $(x_{\mathrm{proj}}, y_{\mathrm{proj}})$. The net line-of-sight velocity, $v_{\mathrm{los}}$, for each galaxy corresponds to the sum of its relative peculiar velocity and the Hubble flow along the line-of-sight axis. This set of three observables are expressed as relative values to the cluster centre. The projected radial distance, $R_{\mathrm{proj}}$, defined as the Euclidean distance from the cluster centre, is derived from the plane-of-sky positions as $R_{\mathrm{proj}} = ( x_{\mathrm{proj}}^2 + y_{\mathrm{proj}}^2 )^{1/2}$. For the computations of $x_{\mathrm{proj}}$, $y_{\mathrm{proj}}$, $R_{\mathrm{proj}}$ and, $v_{\mathrm{los}}$ for a given cluster-galaxy pair, we refer the interested reader to Appendix~A in \citet{ho2019robust}.

\medskip
In this work, we will employ the set of $( R_{\mathrm{proj}}, v_{\mathrm{los}} )$ observables, which define the dynamical phase-space distribution of a galaxy cluster. An example is displayed in the top panel of Fig.~\ref{fig:phase_space_distribution}, which also shows some interloper galaxies. Interlopers are non-member galaxies positioned along the line of sight, with similar observed line-of-sight velocities to the host cluster. The contamination induced by such interlopers is one of the main difficulties involved in cluster mass estimation \citep[e.g.][]{wojtak2007interloper}. Since the clusters will have a varying number of galaxy members or interlopers, we preprocess the mock cluster catalogue, generated as outlined in Section~\ref{mock_catalogues}, by computing the 2D Gaussian kernel density estimate (KDE) of the joint phase-space distribution of $\{ \myvec{R}_{\mathrm{proj}}, \myvec{v}_{\mathrm{los}} \}$, as illustrated in the bottom panel of Fig.~\ref{fig:phase_space_distribution}. A brief introduction to Gaussian KDE is provided in Appendix~\ref{appendix_KDE}. This 2D Gaussian KDE representation subsequently serves as inputs to our neural network (cf. details of architecture and training in Section~\ref{neural_flows}). The extents of the 2D distribution are as follows: $R_{\mathrm{proj}} \in [0, 1.6] \: h^{-1} \, \mathrm{Mpc}$, $v_{\mathrm{los}} \in [-2200, 2200] \: \mathrm{km} \, \mathrm{s}^{-1}$, with 50 bins along a given axis.

\subsection{Mock cluster catalogues}
\label{mock_catalogues}

\begin{figure}
	\centering
		{\includegraphics[width=\hsize,clip=true]{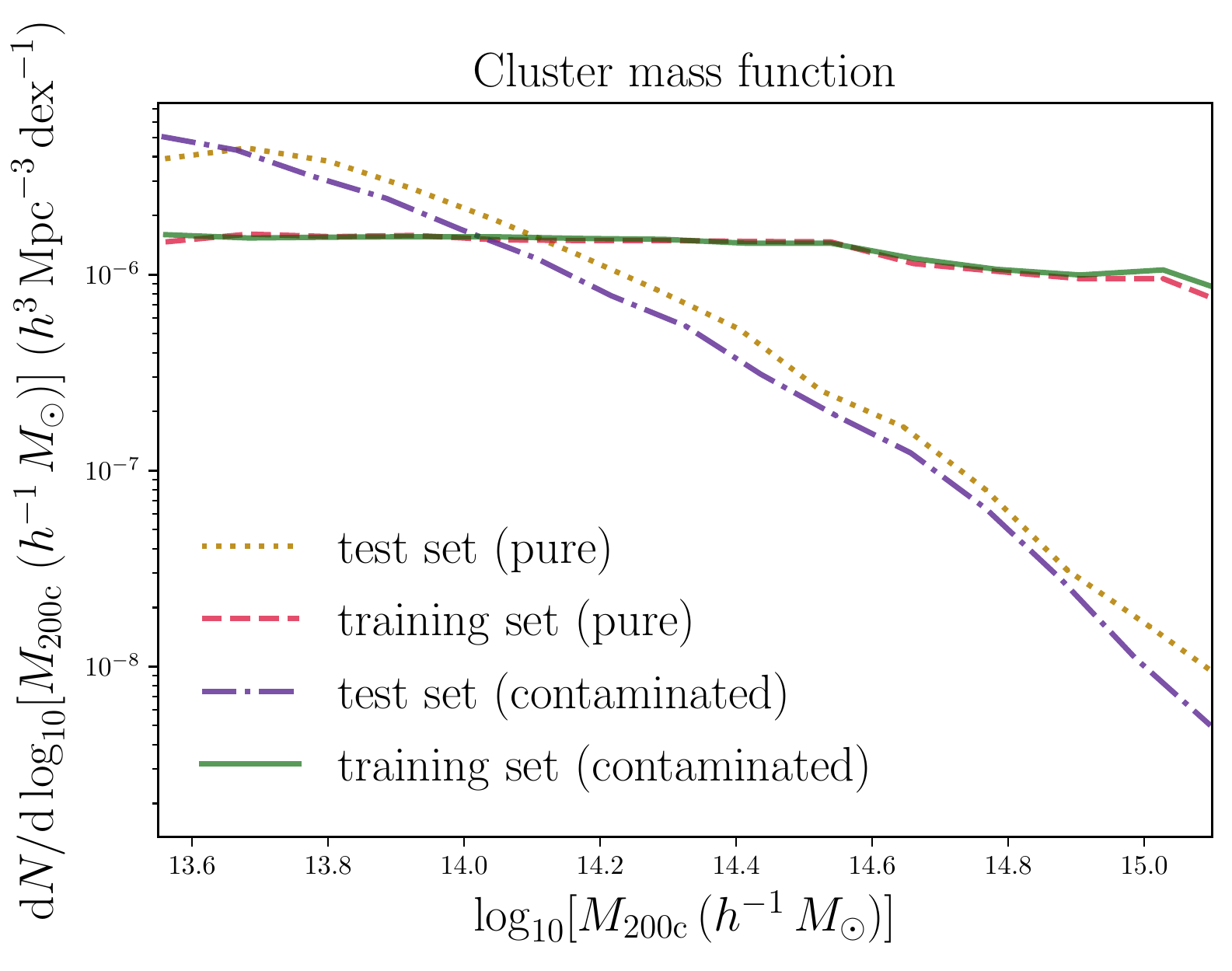}}
	\caption{Mock cluster mass function, i.e. variation of number density of clusters with logarithmic mass for the training and test sets, extracted from the pure and contaminated catalogues, respectively. In order to circumvent any selection bias during training, low-mass clusters are downsampled and high-mass clusters are upsampled along independent lines of sight to produce a flat number density for the training set. The test samples are random cluster subsamples drawn from the actual cosmological mass function. Note that the training and test sets do not contain different line-of-sight augmented versions of the same clusters.}
	\label{fig:cluster_mass_function}
\end{figure}

The mock cluster catalogues are derived from a snapshot ($z = 0.117$) of the MultiDark Planck2 (MDPL2) $N$-body simulation\footnote{\url{https://www.cosmosim.org}} \citep{klypin2016multidark}. MDPL2 is a large dark matter simulation, tracing $3840^3$ particles from an initial redshift $z=120$ to $z=0$, in a cosmological volume of (1 $h^{-1}$ Gpc)$^3$ and mass resolution of $1.51 \times 10^9$ $h^{-1} M_{\odot}$, carried out using \textsc{gadget2} \citep{springel2005gadget2}. The underlying cosmology is set to Planck $\Lambda$CDM best-fit values \citep{planck2014cosmo}: $\Omega_\text{m}=0.307$, $\Omega_{\Lambda}=0.693$, $h=0.678$, $\sigma_8=0.8228$, $n_{\mathrm{s}}=0.96$.

\medskip
A detailed description of the mock generation procedure is provided in Section~2 in \citet{ho2019robust}. The main steps are summarized as follows:
\begin{itemize}
    \item Clusters and their galaxy members are modelled as host halos and subhalos, respectively, which are identified in the MDPL2 simulation using the \textsc{rockstar} halo finder \citep{behroozi2013rockstar}. Clusters are assigned to host halos above a certain mass threshold ($M_{\mathrm{200c}} \geq 10^{13.5}$ $h^{-1} M_{\odot}$), thereby defining their physical properties such as mass, radius, position and velocity. Galaxies, in turn, are associated to subhalos via the galaxy assignment scheme of \textsc{UniverseMachine} \citep{behroozi2019universemachine}, with a selection criterion on their mass at accretion ($M_{\mathrm{acc}} \geq 10^{11}$ $h^{-1} M_{\odot}$), inheriting their corresponding positions and velocities.
    
    \medskip
    \item The mock cluster observations incorporate realistic systematic effects typical of dynamical mass measurements. This includes physical effects such as cluster mergers and triaxiality in the distributions of cluster members, and selection effects, such as the presence of interloper galaxies. Modelling such effects requires selecting member galaxies by extracting a fixed cylindrical volume positioned at the centre of the cluster, aligned with the line-of-sight axis. This is usually referred to as an observational cut.
    
    \medskip
    \item The mock observations do not account for the distance-dependent completeness expected for flux-limited selection of spectroscopic targets. It is assumed that all redshifts above a flux limit corresponding to the adopted minimum subhalo mass are observed.
    
    \medskip
    \item Given the above line-of-sight axis, cluster membership may be determined by computing the position $(x_{\mathrm{proj}}, y_{\mathrm{proj}})$ and velocity ($v_{\mathrm{los}}$) observables, introduced in Section~\ref{dynamical_phase_space_distribution}, for each cluster-galaxy pair. The cylindrical cut is specified in terms of its radial aperture size and half of its length along the line-of-sight axis, with their adopted values being $R_{\mathrm{aperture}} = 1.6 \: h^{-1} \, \mathrm{Mpc}$ and $v_{\mathrm{cut}} = 2500 \: \mathrm{km} \, \mathrm{s}^{-1}$, respectively.
    
    \medskip
    \item Two catalogues are produced from the MDPL2 simulation: {\it pure} and {\it contaminated}, with the primary difference being the presence of interlopers in the latter more realistic version. For the pure catalogue, galaxies located within the virial radius of a particular cluster are included in the mock observation. For the contaminated version, galaxies falling within the cylindrical cut are associated to the host cluster, irrespective of whether they are gravitationally bound to the system. For both catalogues, as a minimum richness criterion, clusters with fewer than ten galaxies are discarded.
\end{itemize}

\medskip
As a three-fold data augmentation procedure, three orthogonal projections are applied to all the original clusters in both the pure and contaminated catalogues. To upsample the number of scarce high-mass clusters in the regime of $M_{\mathrm{200c}} > 10^{14.6} \; h^{-1} M_{\odot}$, additional line-of-sight projections, distributed with roughly even spacing on the unit sphere, are applied. This further marginally augments the size of the catalogues by around $1.5\%$. While the three orthogonal projections are approximately independent, there is a caveat related to the additional projections which introduce some correlations between the augmented high-mass clusters. To verify the performance of our network in an unbiased way, we ensure that the training and test sets do not contain different line-of-sight projections of the same cluster. We, therefore, use the original cluster identifiers from the simulation, as provided by the halo finder, in the assignment procedure. 

\medskip
The cluster mass function, describing the abundance of clusters as a function of mass, of the respective training and test sets, extracted from the simulated pure and contaminated catalogues, is illustrated in Fig.~\ref{fig:cluster_mass_function}. In order not to induce a selection bias during training, we generate a training data set with a flat mass distribution by downsampling the clusters at low masses, which are the most abundant relative to the scarce most massive ones. This may be interpreted as a maximally agnostic prior so as not to encode any cosmological information in our neural network. The test sets, however, follow the theoretical halo mass function that was used in the mock generation to provide a realistic setting to evaluate the performance of our neural network.

\medskip
Both the pure and contaminated training sets contain around twenty thousand clusters randomly drawn from their respective mock catalogues, while ensuring that these two training sets have a flat mass distribution. We designate the corresponding validation sets as $10\%$ of the training sets, such that they contain roughly two thousand clusters each and only eighteen thousand clusters are actually used for training the neural networks. The two test sets, consisting of twenty thousand clusters each, are obtained by randomly sampling from the remaining clusters in the pure and contaminated catalogues, while ensuring that they do not contain augmented versions of the same clusters as in the training sets.

\section{Neural density estimators}
\label{neural_density_estimators}

A neural network, in essence, is a trainable and flexible approximation of a model, $\mathcal{M}(\myvec{\theta}, \myvec{\gamma}): \myvec{d} \rightarrow \myvec{\tau}$, to map some input data $\myvec{d}$ to an estimate of the desired label or target $\myvec{\tau}$ associated with the data. It is parameterized by a set of weights $\myvec{\theta}$, which are tuned via stochastic gradient descent to optimize a given cost or loss function, and a set of hyperparameters $\myvec{\gamma}$, which comprise the choice of network architecture, initialization of the weights, type of activation and loss functions.

\medskip
Density estimation, in its simplest form, entails the problem of estimating the joint probability density $\mathcal{P}(\myvec{x})$ of a set of variables $\myvec{x}$ from a set of examples $\{ \myvec{x}_i \}$. This joint density is crucial to perform a variety of tasks, such as prediction, inference and data generation, and therefore, constitutes a key aspect of probabilistic unsupervised learning and generative modelling. The use of neural networks for density estimation is becoming increasingly popular by virtue of their flexibility and learning capacity and this has led to the emergence of the so-called {\it neural density estimators}. These tools have been particularly successful at modelling natural images \citep[e.g.][]{dinh2016NVP, salimans2017pixelCNN, oord2016conditional} and audio data sets \citep{oord2016wavenet}.

\subsection{Conditional density estimation}
\label{conditional_density_estimation}

Neural density estimators provide a flexible parametric framework to model conditional probability densities, $\mathcal{P}(\myvec{x} | \myvec{y} ; \myvec{\theta})$,\footnote{For simplicity of notation, henceforth, we do not explicitly indicate dependence on the set of hyperparameters $\myvec{\gamma}$.} where $\myvec{\theta}$ corresponds to the weights of the neural network, trained on a set of simulated data(-parameter) pairs $\{ \myvec{x}, \myvec{y} \}$. Unlike other approaches to generative modelling, such as variational autoencoders \citep{kingma2013autoencoding} and generative adversarial networks \citep{goodfellow2014generative}, neural density estimators readily provide exact density evaluations. Ideally, such an estimator should be sufficiently flexible to represent complex distributions and straightforward to train, while having a tractable density function. There are two main classes of neural density estimators which satisfy these conditions: {\it autoregressive models} \citep{uria2016neural} and {\it normalizing flows} \citep{rezende2015normalizing}.

\medskip
The latest advances in deep learning have resulted in the emergence of a particular class of techniques, the so-called neural autoregressive flows, combining normalizing flows and autoregressive models \citep[e.g.][]{kingma2016IAF, papamakarios2017MAF, trippe2018conditional, huang2018NAF}, which have produced state-of-the-art performance in density estimation. We intend to explore a potential extension of our work, as outlined in Section~\ref{conclusions}, to encode such autoregressive flows in a future investigation.

\medskip
This work primarily deals with a conditional density estimation problem, which has garnered significant interest in the field of deep learning. Conditional density estimation, in contrast to the above, is a supervised learning problem, encompassing both classification and regression, where the underlying aim is to predict the distribution of a set of labels $\myvec{x}$ conditional on observing an associated set of features $\myvec{y}$. Here, we employ normalizing flows for conditional density estimation for dynamical mass inference, which are reviewed in Section~\ref{normalizing_flows}.

\medskip
Given their very recent conception and subsequent development in the deep learning community, the use of neural flows for potential astrophysical and cosmological applications remain as yet relatively unexploited. \citet{cranmer2019modeling} employed normalizing flows to model the probability distribution in stellar colour-magnitude space from parallax and photometry measurements. Neural autoregressive flows encoded in likelihood-free inference framework have been used to derive cosmological constraints from supernovae data \citep{alsing2019fast} and to constrain binary black hole systems using gravitational wave data \citep{green2020gravitational}. Such neural flows have also been used within a Bayesian hierarchical model for gravitational wave population inference \citep{wong2020gravitational}.

\subsection{Normalizing flows}
\label{normalizing_flows}

A normalizing flow encodes a smooth invertible mapping between probability density functions via a differentiable, monotonic bijection between the two spaces in which the functions live, i.e. $\mathbb{R}^n \rightarrow \mathbb{R}^n$ \citep{tabak2013family}. The two fundamental requirements are that the transformation must be invertible and that the associated Jacobian must be tractable. The key property of a normalizing flow is that it is composable, such that the composition of a series of relatively simple invertible transformations, applied to a given base distribution, may sufficiently characterize more complex distributions. \citet{rezende2015normalizing} demonstrated that a series of certain families of parametric transformations may warp a standard Gaussian base distribution into a relatively complex target density.

\medskip
A normalizing flow (NoF), as introduced above, characterizes $\mathcal{P}(\myvec{x})$ as an invertible differentiable transformation $\mathcal{F}$ of a base distribution $\Psi (\myvec{u})$, such that $\myvec{x} = \mathcal{F} (\myvec{u})$, where $\myvec{u} \sim \Psi (\myvec{u})$. The rationale behind the choice of the base distribution is that it should be easily evaluated for any input $\myvec{u}$ and hence, it is typically chosen to be a standard Gaussian distribution.\footnote{The base distribution may be any continuous function, such as a mixture of Gaussians, which would be more flexible than a single Gaussian distribution.} Assuming that the transformation $\mathcal{F}$ is invertible, we can compute $\mathcal{P}(\myvec{x})$ via a change of variables as
\begin{equation}
    \mathcal{P}(\myvec{x}) = \Psi \big[ \mathcal{F}^{-1} (\myvec{x}) \big] \left|  \frac{\partial \mathcal{F}^{-1} (\myvec{x})}{\partial \myvec{x}} \right| ,
    \label{eq:normalizing_flow}
\end{equation}
which implies that $\mathcal{F}$ must be invertible and that it should be possible to compute the determinant of its Jacobian for equation~\eqref{eq:normalizing_flow} to be tractable, substantiating the two key requirements laid out above. An interesting point is that if two transformations, $\mathcal{F}_1$ and $\mathcal{F}_2$, satisfy both conditions, then so does the composition $\mathcal{F}_1 \circ \mathcal{F}_2$. As such, subsequently stacking multiple instances of $\mathcal{F}$ would still preserve the properties of NoF, while resulting in a deeper and more flexible model.

\medskip
The invertibility of $\mathcal{F}$ implies that a NoF allows both sampling and probability density evaluation of $\mathcal{P}(\myvec{x})$, as long as these tasks are feasible for the base distribution $\Psi (\myvec{u})$. To sample from $\mathcal{P}(\myvec{x})$, we first draw samples $\myvec{u} \sim \Psi(\myvec{u})$ to then obtain the desired samples via $\myvec{x} = \mathcal{F}(\myvec{u})$. For density evaluation, we must inject the inverse mapping $\myvec{u} = \mathcal{F}^{-1} (\myvec{x})$ in equation~\eqref{eq:normalizing_flow}.

\subsection{Conditional neural flows}
\label{conditional_neural_flows}

Conditional density estimation, in essence, involves estimating the conditional density distribution $\mathcal{P}(\myvec{x}|\myvec{y})$ from a given set of data pairs $\{ \myvec{x}_n, \myvec{y}_n \}$. The unconditional NoF framework may be adapted to parameterize a conditional distribution simply by adding $\myvec{y}$ to the set of input variables. Our choice of NoF design is inspired by the Masked Autoencoder for Distribution Estimation\footnote{Our neural network architecture reduces to a particular implementation of NoFs inspired by the masked autoregressive flow \citep[MAF,][]{papamakarios2017MAF} which involves a stack of several MADEs. In our special variant of MAF, we discard the binary mask in the MADE block, introduced to preserve the autoregressive (chain rule of probability) property, since we work in terms of a single autoregressive conditional distribution (due to cluster mass $M$ being a scalar).} \citep[MADE,][]{germain2015made}, which is typically used in autoregressive models.

\medskip
The NoF framework, in a nutshell, learns the means and variances of the conditional distributions. In terms of architecture, the intermediate or hidden layers of one NoF block may encode non-linear activation functions (e.g. \texttt{tanh}, \texttt{sigmoid}, \texttt{ReLU}). The output nodes, however, must obey certain constraints, with the nodes corresponding to the conditional means having a linear activation, whilst those associated with conditional variances must have exponential activations to ensure positivity. Our neural flow model is a stack of multiple NoF blocks of the same type to yield an overall NoF with higher flexibility than the original one, i.e. $\mathcal{F} \equiv \mathcal{F}_1 \circ \mathcal{F}_2 \circ \ldots \mathcal{F}_k$ for $k$\textsuperscript{th} component in the neural flow. The output of each NoF block is fed to the next one, along with the conditional inputs (cf. Fig.~\ref{fig:neural_flow_schematic}). Since density evaluations are possible from equation~\eqref{eq:normalizing_flow}, the neural network parameterizing the overall NoF may be trained using stochastic gradient descent to maximize the likelihood over the set of network weights $\myvec{\theta}$ that the data pairs $\{ \myvec{x}_n, \myvec{y}_n \}$ emanate from the model.

\medskip
Our implementation reduces to a simplified version, since $\mathrm{dim}(\myvec{x}) = 1$ in our case, since we are inferring only the dynamical cluster mass $M$ from the phase-space distribution $\{ \myvec{R}_{\mathrm{proj}}, \myvec{v}_{\mathrm{los}} \} \equiv \tilde{\myvec{d}}$, such that a given NoF block must learn solely the mean and variance of a single conditional (Gaussian) distribution. The conditional NoF may be interpreted as learning the transformation of the random variate $M$ to the latent space $u$ where we set the unit normal distribution, $u (M, \tilde{\myvec{d}} ; \myvec{\theta}) \sim \mathcal{N}(0, 1)$, via
\begin{equation}
	u = \frac{M - \mu_*(\tilde{\myvec{d}} ; \myvec{\theta}) }{\sigma_*(\tilde{\myvec{d}} ; \myvec{\theta}) } ,
    \label{eq:conditional_NoF_transform}
\end{equation}
where $\mu_*$ and $\sigma_*$ correspond to the mean and variance, respectively, of the 1D conditional distribution, with $\mu_* \in \mathbb{R}^n, \; \sigma_* \in \mathbb{R}^n_{+}$ , and $\myvec{\theta}$ are the trainable weights of the neural network. The Jacobian of the above invertible mapping is, therefore, trivial, such that we can express the conditional density estimator of a given NoF component, from equations~\eqref{eq:normalizing_flow} and \eqref{eq:conditional_NoF_transform}, as
\begin{align*}
    \mathcal{P}(M | \tilde{\myvec{d}}; \myvec{\theta}) &= \mathcal{N} \big[ u (M, \tilde{\myvec{d}} ; \myvec{\theta}) \big] \times \left| \frac{\partial u (M, \tilde{\myvec{d}} ; \myvec{\theta}))}{\partial M} \right| \\
    &= \mathcal{N} \big[ u (M, \tilde{\myvec{d}} ; \myvec{\theta}) \big] \times \sigma_*(\tilde{\myvec{d}} ; \myvec{\theta})^{-1} . \numberthis
    \label{eq:conditional_NoF_estimator}
\end{align*}

\medskip
Driven by the above rationale, we can implement the conditional neural flow model by stacking multiple such conditional NoFs, i.e. there is now an additional input $\tilde{\myvec{d}}$ to every layer (cf. Fig.~\ref{fig:neural_flow_schematic}). From a straightforward extension of the NoF conditional density estimator given by equation~\eqref{eq:conditional_NoF_estimator}, we can express the conditional density estimator characterized by the overall neural flow model as
\begin{equation}
    \mathcal{P}(M | \tilde{\myvec{d}}; \myvec{\theta}) = \mathcal{N} \big[ u_{\mathrm{out}} (M, \tilde{\myvec{d}} ; \myvec{\theta}) \big] \times \prod^{N_{\textsc{n}\mathrm{o}\textsc{f}}}_{k=1} \sigma^k_*(\tilde{\myvec{d}} ; \myvec{\theta})^{-1} ,
    \label{eq:conditional_neural_flow_estimator}
\end{equation}
where $N_{\textsc{n}\mathrm{o}\textsc{f}}$ denotes the number of NoF blocks in the overall neural flow, with $k$ labelling each NoF component, and $u_{\mathrm{out}}$ corresponds to the output from the final NoF block.

\section{Neural flows}
\label{neural_flows}

In this section, we describe our implementation of neural flows, the network architecture and training rationale.

\subsection{Neural network architecture}
\label{neural_network_architecture}

\begin{figure*}
	\centering
		{\includegraphics[width=\hsize,clip=true]{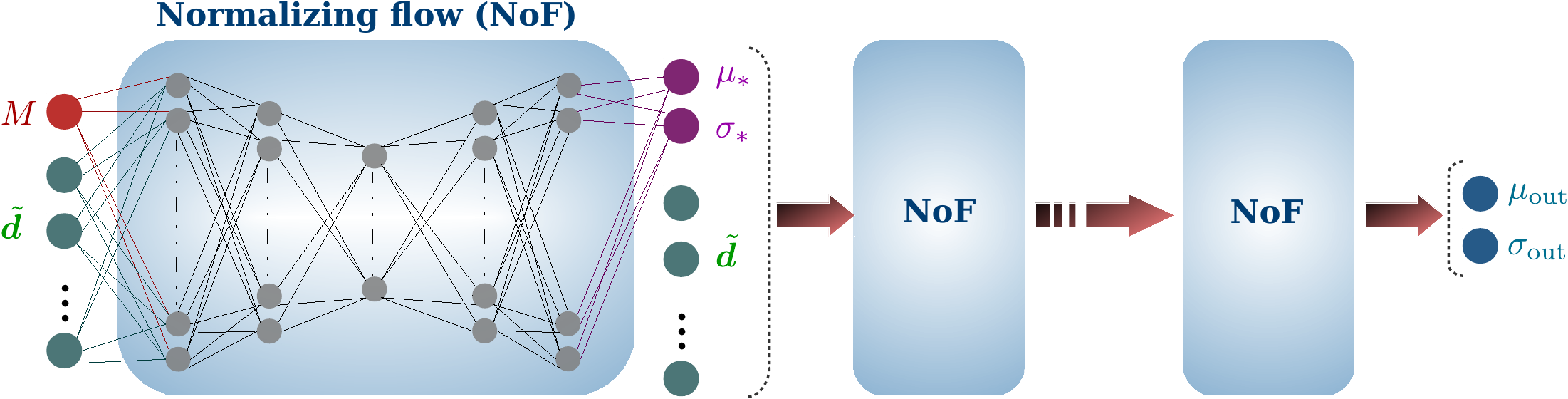}}
	\caption{Schematic representation of our neural flow architecture to model the conditional density distribution $\mathcal{P}(M | \tilde{\myvec{d}})$. The detailed architecture of only one normalizing flow (NoF) component is displayed, with the neural flow being composed of a series of such blocks. The first NoF component takes as input the cluster mass $M$ and the corresponding 2D KDE of the phase-space distribution $\tilde{\myvec{d}}$ to yield the conditional mean and variance, $\mu_*$ and $\sigma_*$, respectively. The output of each NoF block is fed as input to the next one, along with the conditional inputs $\tilde{\myvec{d}}$. The network is trained using pairs of $\{ M, \tilde{\myvec{d}} \}$ from the training set by minimizing the negative log-likelihood from equation~\eqref{eq:loss_conditional_special_NoF}. Once optimized, the cluster mass PDF $\mathcal{P}(M | \tilde{\myvec{d}})$ may be computed by injecting the output of the final NoF block, computed as $u_{\mathrm{out}} = (M - \mu_{\mathrm{out}})/\sigma_{\mathrm{out}}$, in equation~\eqref{eq:conditional_neural_flow_estimator} for a range of cluster masses $M$, along with the $\sigma_*$'s from all the NoF components.} 
	\label{fig:neural_flow_schematic}
\end{figure*}

The underlying objective of our neural flow framework is to model the conditional probability density function (PDF) of the dynamical mass of a galaxy cluster, given its phase-space distribution characterized by the projected radial distance from cluster centre and the line-of-sight velocity, i.e. $\mathcal{P}(M | \{ \myvec{R}_{\mathrm{proj}}, \myvec{v}_{\mathrm{los}} \})$. The network takes as input a vector $\tilde{\myvec{d}}$, which is a flattened 2D KDE of the phase-space distribution, i.e. $\tilde{\myvec{d}} \equiv \{ \myvec{R}_{\mathrm{proj}}, \myvec{v}_{\mathrm{los}} \}$, with an example illustrated in Fig.~\ref{fig:phase_space_distribution}. The neural flow, therefore, models $\mathcal{P}(M | \tilde{\myvec{d}})$, with the training data set consisting of pairs of $\{ M, \tilde{\myvec{d}} \}$.

\medskip
A general schematic of our neural flow architecture is depicted in Fig.~\ref{fig:neural_flow_schematic}. We employ three NoF components, with a Gaussian base distribution and \texttt{tanh} activation in the three hidden layers of each component, with 512, 64 and 512 neurons, respectively. The hidden layers of each NoF block are standard fully-connected layers, with a given pair of $\{ M, \tilde{\myvec{d}} \}$ fed to the first NoF component, assuming a batch size of unity for simplicity. The outputs of the intermediate NoF blocks are the conditional mean and variance, $\mu_*$ and $\sigma_*$, respectively, which are then used to compute $u_*$ via equation~\eqref{eq:conditional_NoF_transform}. The latter quantity is subsequently fed to the next NoF component, along with the conditional input $\tilde{\myvec{d}}$. This procedure is repeated for all intermediate NoF blocks until the final one, which computes $u_{\mathrm{out}}$ by plugging the outputs $\mu_{\mathrm{out}}$ and $\sigma_{\mathrm{out}}$ in equation~\eqref{eq:conditional_NoF_transform}. The same steps apply for any arbitrary batch size. The procedure to obtain the conditional PDF of the mass of a given cluster using the above architecture is described in the next Section. As explained in Section~\ref{conditional_neural_flows}, the output nodes associated to the conditional mean and variance must, respectively, have linear and exponential activations to guarantee positivity.

\subsection{Training methodology}
\label{training_methodology}

In order to fit a neural density estimator to our set of data pairs $\{ M, \tilde{\myvec{d}} \}$, we must optimize the weights of the neural network to minimize the Kullback-Leibler (KL) divergence between the parametric neural density estimator, $\mathcal{P}( M | \tilde{\myvec{d}}; \myvec{\theta})$, and the target distribution, $\hat{\mathcal{P}}( M | \tilde{\myvec{d}})$ \citep[cf.][for more details]{alsing2019fast}, as the loss function. A Monte-Carlo estimate of this KL divergence yields the negative log loss function as a sum over the samples in our training data set, as follows:
\begin{equation}
    - \ln \mathcal{L} (\myvec{\theta}| \{ M, \tilde{\myvec{d}} \}) = - \sum_{i=1}^{N_{\mathrm{samples}}} \ln \mathcal{P}(M | \tilde{\myvec{d}}; \myvec{\theta}) ,
    \label{eq:loss_NDE}
\end{equation}
where $N_{\mathrm{samples}}$ is the size of our training set. Note that this is equivalent to the negative log-likelihood of the simulated data $\{ M, \tilde{\myvec{d}} \}$ under the conditional density estimator $\mathcal{P}( M | \tilde{\myvec{d}}; \myvec{\theta})$. In the case of a conditional NoF, with a Gaussian base distribution as given by equation~\eqref{eq:conditional_neural_flow_estimator}, we may express equation~\eqref{eq:loss_NDE} as
\begin{multline}
    - \ln \mathcal{L} (\myvec{\theta}| \{ M, \tilde{\myvec{d}} \}) = - \sum_{i=1}^{N_{\mathrm{samples}}} \ln \mathcal{N} \big[ u_{\mathrm{out}} ( M_i, \tilde{\myvec{d}}_i ; \myvec{\theta}) \big] + \\ \sum_{k=1}^{N_{\textsc{n}\mathrm{o}\textsc{f}}} \ln \sigma^k_* (M_i, \tilde{\myvec{d}}_i ; \myvec{\theta}) ,
    \label{eq:loss_conditional_special_NoF}
\end{multline}
where $u_{\mathrm{out}}$ is the output of the final component, $u_{\mathrm{out}} = (M - \mu_{\mathrm{out}})/\sigma_{\mathrm{out}}$.

\medskip
We train our neural flow model by minimizing the negative log loss function given by equation~\eqref{eq:loss_conditional_special_NoF}, with respect to the network weights $\myvec{\theta} = \{ \myvec{\theta}_j  \}$ for the $j$\textsuperscript{th} hidden layer. In order to obviate risks of overfitting, we adopt a standard regularization method of early stopping in our training routine. To this end, we split the mock cluster catalogue into a training and validation data set. We designate an early stopping criterion of 200 weight updates, such that training is terminated when the validation loss no longer shows any improvement for this chosen number of consecutive training iterations, and the previously saved best fit model is restored.

\medskip
The neural network and training procedure are implemented in \textsc{TensorFlow} \citep{abadi2016tensorflow}. We use the {\it Adam} \citep{kingma2014adam} optimizer, with a learning rate of $\eta=10^{-4}$ and first and second moment exponential decay rates of $\beta_1=0.9$ and $\beta_2=0.999$, respectively. The batch size is set to 100. We train the network for $\sim\, 2\times10^3$ weight updates (i.e. training iterations over a given batch), requiring around five minutes on an NVIDIA V100 Tensor Core GPU. We train two different networks, but with the same architecture, on the respective pure and contaminated catalogues. The pure catalogue provides the ideal scenario without the presence of interlopers, the main contaminant in the mass determination. Comparing the performance of the model in the ideal and realistic settings would provide some insights pertaining to the effectiveness of the neural flow to account for the spurious contaminations.

\medskip
Once the model is optimized, in order to obtain the conditional PDF $\mathcal{P}(M | \tilde{\myvec{d}})$ of the mass of a given cluster, the output of the final component of the neural flow model, computed as $u_{\mathrm{out}} = (M - \mu_{\mathrm{out}})/\sigma_{\mathrm{out}}$, must be injected in equation~\eqref{eq:conditional_neural_flow_estimator} for a range of cluster masses $M$, along with the $\sigma_*$'s from all the NoF blocks. In other words, the network predictions for a given $\tilde{\myvec{d}}$ of a particular cluster across a range of masses allow us to compute the mass PDF for the cluster of interest.

\section{Validation and performance}
\label{validation_performance}

\begin{figure*}
	\centering
    \subfloat[Pure catalogue]{\includegraphics[width=0.475\hsize]{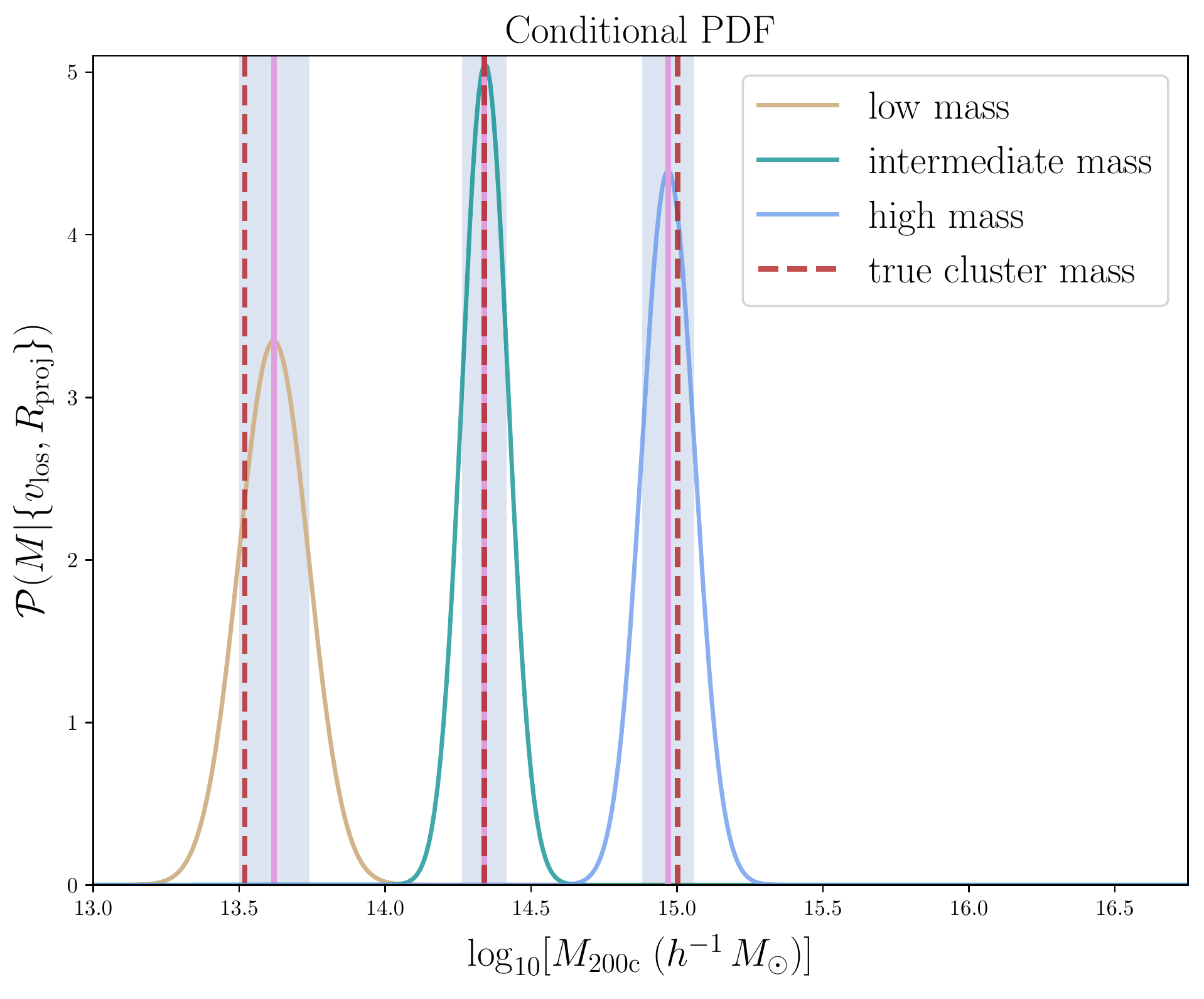}}
    \quad
    \subfloat[Contaminated catalogue]{\includegraphics[width=0.475\hsize]{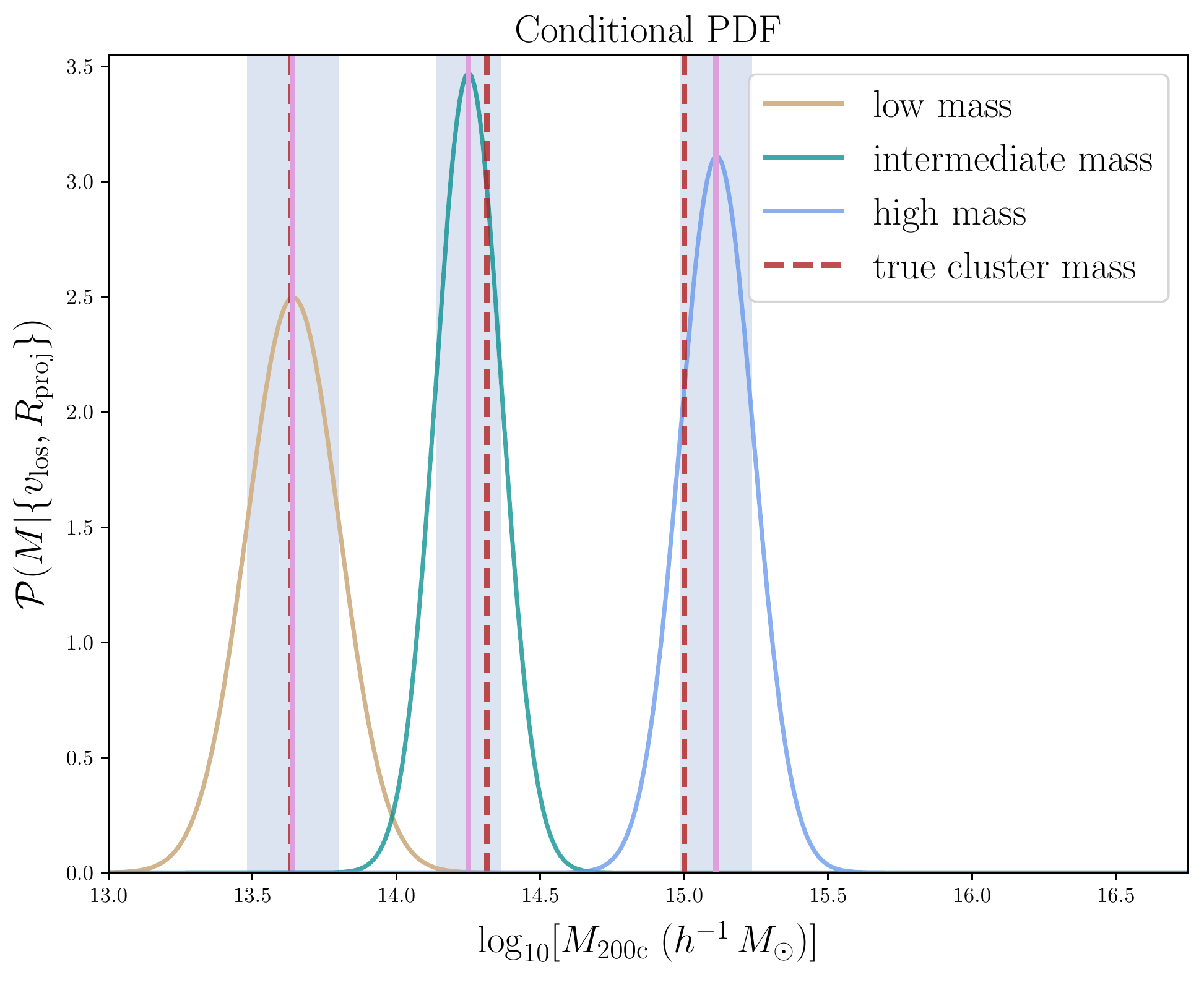}}
    \quad
	\caption{Model predictions for three individual galaxy clusters in distinct regimes of the cluster mass range of the simulated (a) pure and (b) contaminated  catalogues. The individual mass PDFs from our NF mass estimator are illustrated, with the solid line and shaded regions depicting their respective means and $1\sigma$ confidence regions. The corresponding ground truth masses are indicated via the dashed lines. As a consistency test, this validates the capability of our neural network to properly recover the actual cluster masses and demonstrates the potential constraining power of our approach, with the $1\sigma$ uncertainties gradually reducing to $\sim 0.08$~dex and $\sim 0.13$~dex, for the most massive clusters in the pure and contaminated scenarios, respectively. As expected, the relatively larger uncertainties for the latter catalogue reflect the contamination by interloper galaxies.}
	\label{fig:individual_cluster_mass_predictions}
\end{figure*}

\begin{figure*}
	\centering
    \subfloat[Pure catalogue]{\includegraphics[width=0.475\hsize]{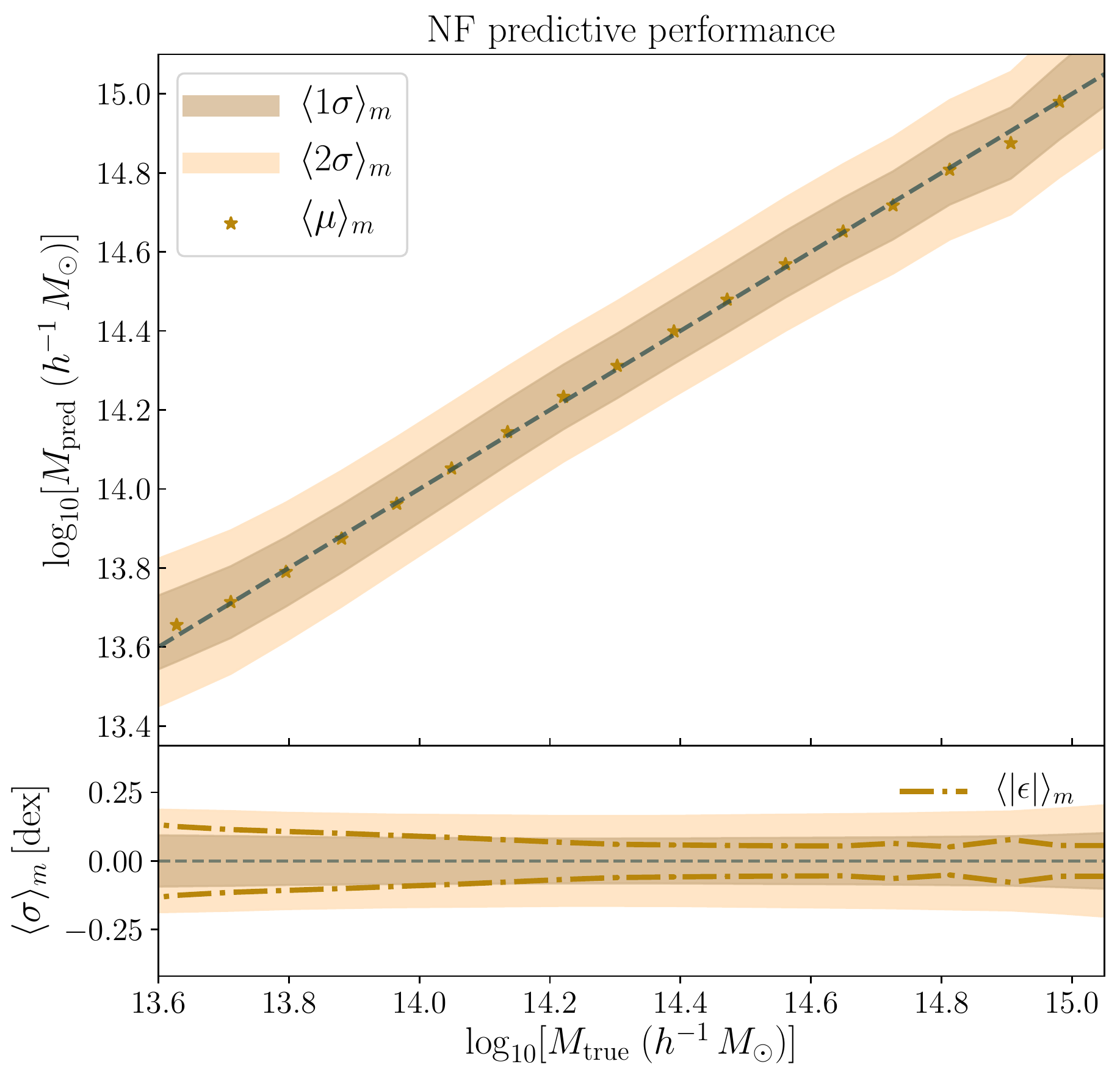}}
    \quad
    \subfloat[Contaminated catalogue]{\includegraphics[width=0.475\hsize]{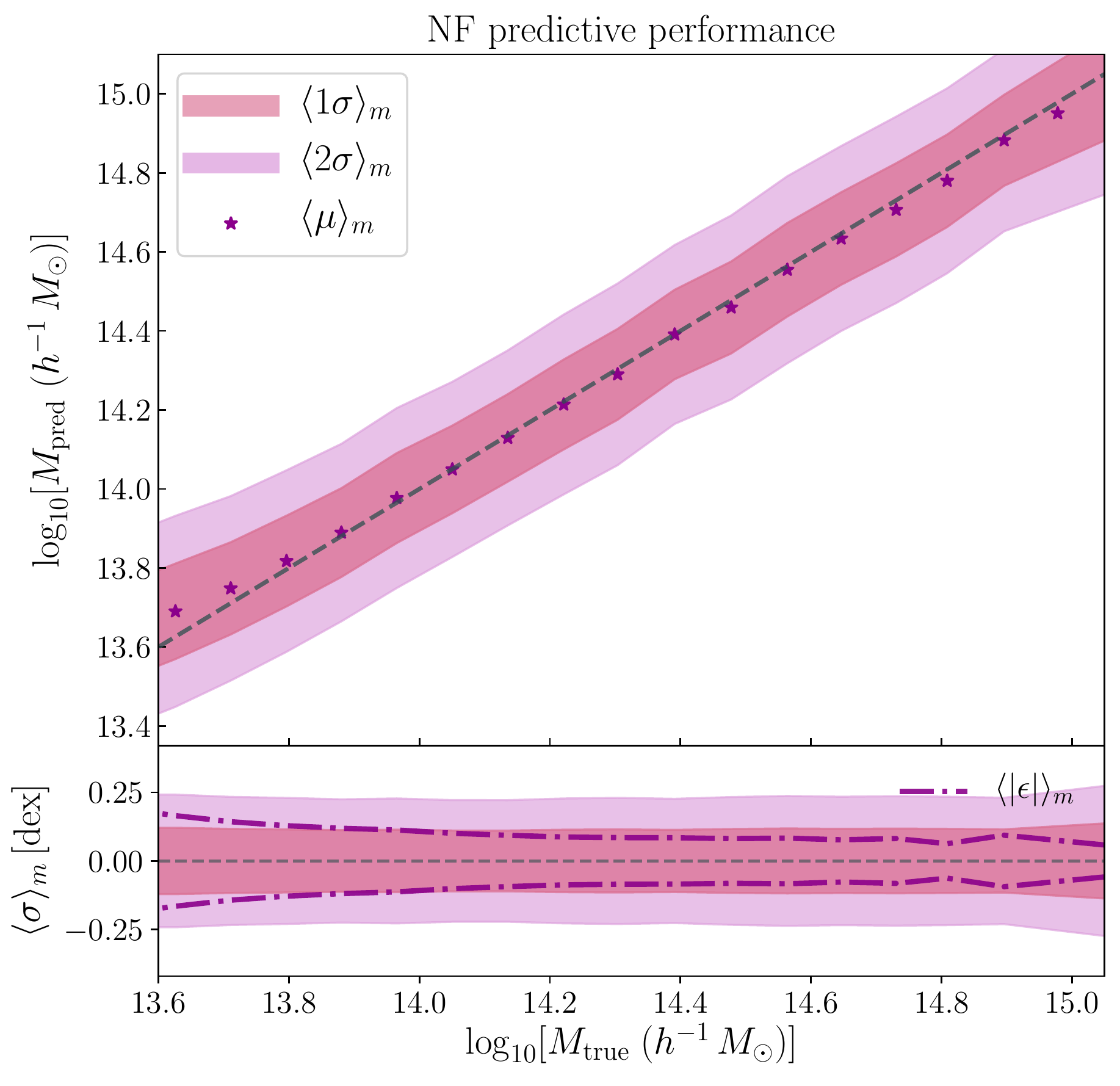}}
    \quad
	\caption{Model predictions against ground truth, showing the means, $1\sigma$ and $2\sigma$ confidence regions, averaged per logarithmic mass bin for the simulated (a) pure and (b) contaminated  catalogues. {\it Top panels:} The coloured stars indicate the respective means, while the $1\sigma$ and $2\sigma$ uncertainties are represented via the shaded bands. {\it Bottom panels:} Variation of the $1\sigma$ and $2\sigma$ uncertainties as a function of logarithmic mass. The relatively larger uncertainties for the contaminated clusters are primarily due to the presence of interlopers. Nevertheless, the mean predictions across the range of mass bins are all within the $1\sigma$ region, thereby illustrating the efficacy and robustness of our neural flow model. If we consider only point predictions from our NF mass estimator, then the dash-dotted lines in the bottom panels would represent the average absolute residual scatter about the ground truth.}
	\label{fig:combined_cluster_mass_predictions}
\end{figure*}

\begin{figure}
    \centering
    \includegraphics[width=\hsize]{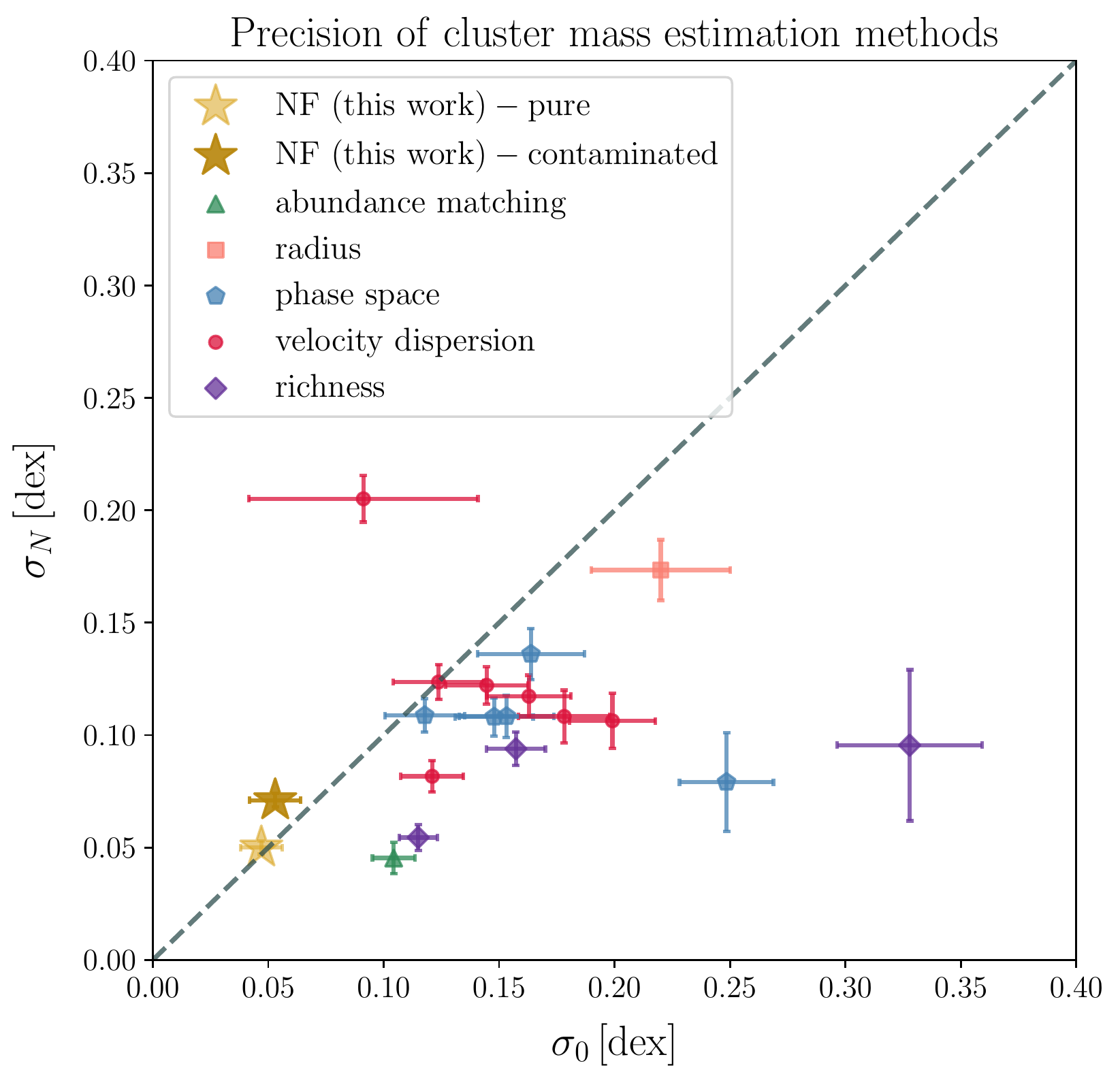}
    \caption{Comparison between the precision of our NF mass estimator and seventeen traditional methods deriving cluster masses by means of measuring richness, velocity dispersion, radial distribution of galaxies, projected phase-space distribution of galaxies and abundance matching \citep{old2015clusterII}. The precision is quantified in terms of richness-dependent error with amplitude $\sigma_{N}$ (error for a cluster with 100 galaxies) and richness-independent systematic error $\sigma_{0}$, as defined by equation~\eqref{eq:errors}. The neural network approach devised in this study, as indicated by the stars, outperforms the traditional techniques.}
    \label{fig:sigma_comparison}
\end{figure}

\medskip
We evaluate the performance of our neural flow mass estimator (hereafter NF mass estimator) on independent test sets extracted from their respective simulated pure and contaminated catalogues described in Section~\ref{mock_catalogues}. The test sets follow the cluster mass function (cf. Fig.~\ref{fig:cluster_mass_function}) employed in the mock generation, unlike the flat mass distribution of the training sets, in order to assess the performance of the model under realistic conditions.

\medskip
Fig.~\ref{fig:individual_cluster_mass_predictions} depicts the model predictions for three particular galaxy clusters of low, intermediate and high masses selected from the mass range of our catalogue. We find that the inferred PDFs have remarkably small $1\sigma$ uncertainties, at the level of $\sim 0.08$~dex and $\sim 0.13$~dex for the most massive clusters in the pure and contaminated catalogues, respectively, and closely match the corresponding ground truth masses. The relatively larger uncertainties for the contaminated clusters result primarily from selection effects such as the presence of interlopers, encountered in practice, as explained in Sections~\ref{classical_m_sigma} and \ref{mock_catalogues}.

\medskip
An essential feature of our approach is that it yields robust uncertainties for individual cluster mass predictions. In contrast, the standard machine learning methods for cluster mass estimation, as reviewed in Section~\ref{intro}, yield point estimates and typically assess the residual scatter, $\epsilon \equiv \log_{10} (M_{\mathrm{true}} / M_{\mathrm{pred}})$, in their predictions relative to the ground truth. In Fig.~\ref{fig:combined_cluster_mass_predictions}, we illustrate the $1\sigma$ and $2\sigma$ confidence regions of our model predictions, against the ground truth masses, for the clusters in the test set across the complete mass range. For the sake of comparison, for instance, with Fig.~7(b) in \citet{ho2019robust}, and for a fair representation of our network performance, the model predictions are binned in logarithmic mass intervals, labelled by $m$, with the respective means and standard deviations, averaged per bin, displayed in the top row of Fig.~\ref{fig:combined_cluster_mass_predictions} via the coloured stars and shaded regions, respectively, for both the pure and contaminated catalogues. The corresponding bottom panels indicate the variation of the average $1\sigma$ and $2\sigma$ uncertainties across the different mass bins.

\begin{figure}
    \includegraphics[width=\hsize]{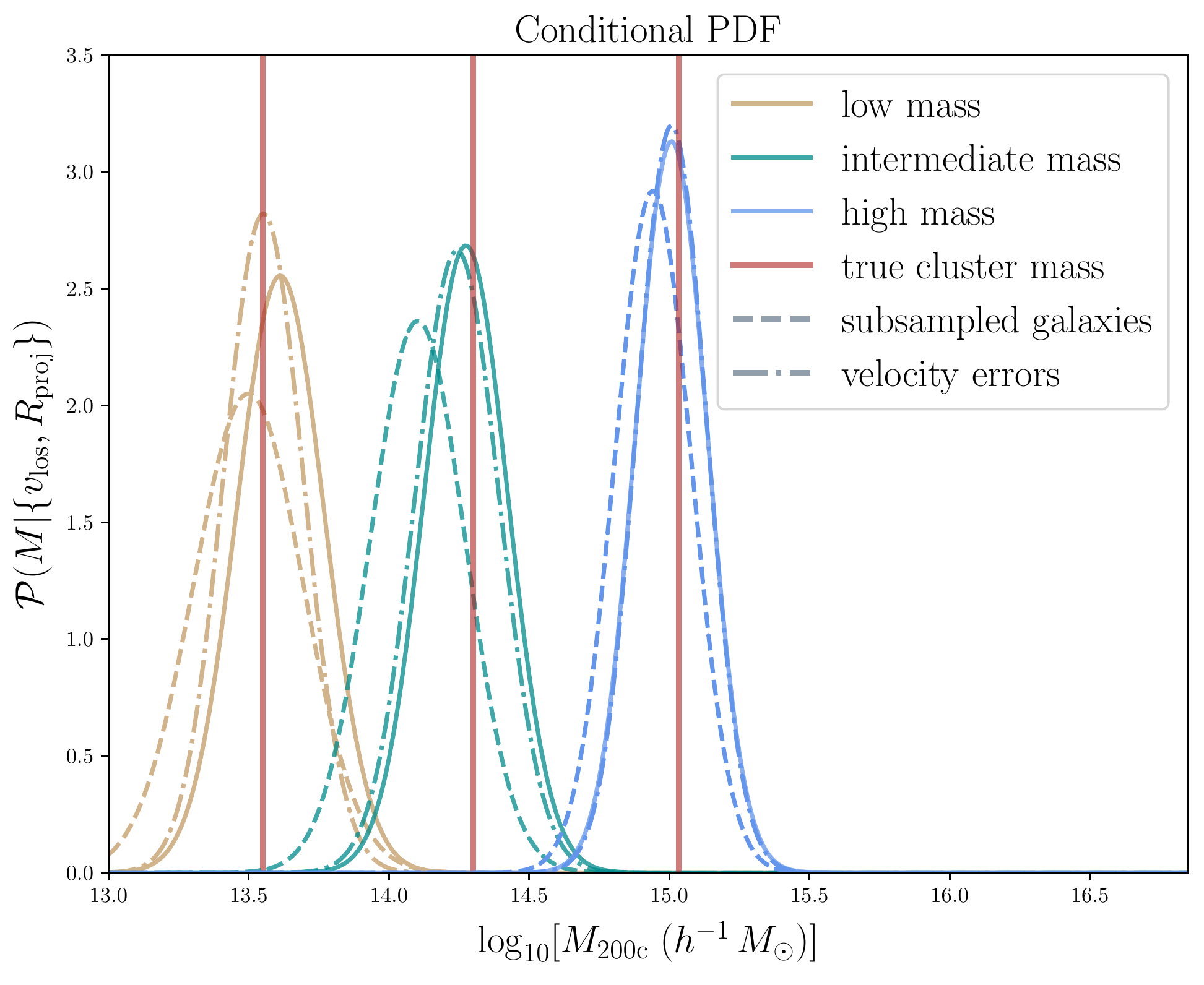}
	\caption{Robustness of our NF mass estimator to the size of galaxy samples with spectroscopic redshifts and velocity errors. The model predictions for three individual galaxy clusters (solid lines) from the contaminated test set, along with the inferred PDFs for these three clusters when randomly subsampled to have $33\%$ fewer galaxies (dashed lines) and with Gaussian scatter of 100~km/s applied to the velocities of the galaxies of the original clusters (dash-dotted lines). These two tests demonstrate the robustness of our neural flow approach to galaxy selection effects and typical velocity errors from current spectroscopic observations.}
	\label{fig:robustness_tests}
\end{figure}

\medskip
If we consider only point predictions from our model, then the absolute logarithmic residual scatter, $\langle |\epsilon| \rangle_m$, in our model predictions would be given by the dash-dotted lines in the bottom panels of Fig.~\ref{fig:combined_cluster_mass_predictions}, which correspond roughly to the $1\sigma$ band depicted. For the contaminated test set with interlopers, the overall logarithmic residual scatter is $\langle \epsilon \rangle = 0.028$~dex with a log-normal scatter of $\sigma_{\epsilon} = 0.126$~dex. In the mass range of $14.0 \leq \log_{10} (M_{\mathrm{true}}) \leq 15.0$~dex, as considered by \citet{ho2019robust}, $\sigma_{\epsilon} = 0.089$~dex, which is an improvement by $\sim 33\%$ relative to that obtained with the 2D CNN in \citet{ho2019robust}. The latter, in turn, is an improvement by a factor of three over the classical $M-\sigma$ power-law estimator. As such, our NF mass estimator yields an overall improvement by nearly a factor of four relative to the classical scaling relation. The average absolute residual scatter $\langle|\epsilon|\rangle$ drops to 0.067~dex for the most massive contaminated clusters in the range of $14.8 \leq \log_{10} (M_{\mathrm{true}}) <15.2$. The test set contains around 50 clusters in this particular mass range, with the most significant outliers being two clusters with $|\epsilon| \simeq 0.24$~dex. For completeness, the pure test set has $\langle \epsilon \rangle = 0.007$~dex, with $\sigma_{\epsilon} = 0.058$~dex. As expected, our neural flow model assigns larger uncertainties to the contaminated samples, but the mean predictions are all within the $1\sigma$ band, further substantiating the reliability and accuracy of our NF mass estimator even when dealing with the presence of interloper galaxies, on top of the physical contaminations which distort the galaxy cluster shape or mass distribution.

\medskip
In Fig.~\ref{fig:sigma_comparison}, we compare the precision of the NF mass estimator and seventeen different methods of cluster mass estimation based on galaxy data. The estimates of precision for the reference methods are from the  Galaxy Cluster Mass Comparison Project, which delivered an extensive test of a wide range of cluster mass estimation techniques using two contrasting mock galaxy catalogues \citep{old2015clusterII}. Cluster masses were estimated using different variations of methods based on measuring the number of galaxy cluster members (richness), the velocity dispersion, the distribution of galaxy positions (projected radii), the galaxy distribution in the projected phase space and abundance matching \citep[see more details in][]{old2015clusterII}. We show the results obtained for the mock galaxy catalogue generated with a semi-analytic model of galaxy formation, which is expected to resemble closely the mock contaminated cluster data used in our study. The precision of cluster mass estimation is quantified in terms of the total scatter about the best-fit power-law relation between the true and estimated cluster masses. Following \citet{wojtak2018clusterIV}, we split the total scatter $\sigma$ into richness-dependent $\sigma_{N}$ and richness-independent $\sigma_{0}$, as given by
\begin{equation}
\sigma^{2} = \sigma_{N}^{2}(N_{\rm mem}/100)^{-1} + \sigma_{0}^{2},
\label{eq:errors}
\end{equation}
where $N_{\rm mem}$ is the number of cluster members. Both $\sigma_{N}$ and $\sigma_{0}$ are determined by fitting the above equation to the logarithmic residuals in the cluster mass measurements from the test runs. If systematic errors are negligible, $\sigma$ becomes a statistical error given by Poisson noise with the amplitude equal to $\sigma_{N}$. In general, systematic effects, depending on whether or not they scale with richness, can increase both $\sigma_{0}$ and $\sigma_{N}$.

\medskip
Fig. 6 demonstrates that the NF mass estimator devised
in our study outperforms the traditional methods. For the contaminated data set, we find $\sigma_{N}=0.07$~dex and $\sigma_{0}=0.05$~dex, which are clearly below typical values attained by the traditional techniques. The richness-dependent error is also smaller than the $0.09$~dex expected for the mass estimation based solely on the scaling relation with the velocity dispersion, i.e. $3/(\sqrt{2N_{\rm mem}}\ln10)$. This shows how much the constraining power increases when the full information on the projected phase-space distribution of galaxies is exploited instead of relying merely on the velocity dispersion. For the pure galaxy catalogue, we find $\sigma_{N}=0.05$~dex and $\sigma_{0}=0.05$~dex. When comparing these parameter values to those obtained for the contaminated catalogue, one can conclude that systematic error caused by interlopers scales with cluster richness and becomes the dominant source of errors for low-richness systems. A similar trend was also shown for more traditional methods of cluster mass estimation \citep{wojtak2018clusterIV}.

\medskip
We perform two tests to demonstrate the robustness of our method to galaxy selection effects, i.e. the size of galaxy samples with spectroscopic redshifts, and velocity errors from spectroscopic observations, respectively. First, to verify the reliability of our model predictions when a fraction of galaxy members of a given cluster are not observable or distinguishable, we randomly subsample three clusters from the contaminated test set in different mass regimes to have $33\%$ fewer galaxies. The inferred mass PDFs are illustrated in Fig.~\ref{fig:robustness_tests} in dashed lines, with the solid lines indicating the predictions for the original clusters with no subsampling. From the inferred PDFs, we find that the subsampled clusters have marginally lower mass estimates, with the effect being larger for the low-mass cluster as expected. In all three cases, the ground truth masses fall within $1\sigma$ of the maximum {\it a posteriori} estimates. In order to demonstrate the robustness to velocity errors typical of current spectroscopic observations, we apply a random scatter, drawn from a zero-centred Gaussian distribution with standard deviation of 100~km/s to the line-of-sight velocities of the galaxies from the above three clusters. This is roughly three times larger than the typical error of spectroscopic redshifts in the SDSS Main Galaxy Sample \citep{Strauss2002}. The corresponding inferred mass PDFs are depicted in dash-dotted lines on Fig.~\ref{fig:robustness_tests}, with no significant bias induced by such velocity errors.

\section{Applications}
\label{applications}

We now apply the neural network trained on the contaminated catalogue to redshift data of several real galaxy clusters to infer their dynamical masses and eventually make a comparison with corresponding masses available from the literature. The main goal is to demonstrate the potential of the NF mass estimator to recover well confirmed dynamical mass measurements obtained for observed nearby galaxy clusters. For this purpose, we select well-studied low-redshift galaxy clusters with large samples of spectroscopically measured redshifts and robust dynamical mass estimates. The selected clusters include the Coma cluster, the lensing cluster A1689 and six rich galaxy clusters for which dynamical models yield consistent mass estimates for a few independent methods of removing relatively strong interloper contamination \citep{Wojtak2007_clusters}.

\begin{table}
    \caption[Dynamical mass inference of real galaxy clusters]{\label{tab:inference_real_clusters} 
    Dynamical masses of selected well-studied galaxy clusters, as predicted by our NF mass estimator, with the measurements available from the literature also indicated. Note that the masses are listed as $\log_{10}[M_{\mathrm{200c}} \, (h^{-1} M_{\odot})]$.}
    \begin{center}
        \begin{threeparttable}
            \begin{tabular}{lll}
            \hline
            Galaxy cluster & NF dynamical mass & Literature value \\
            \hline \hline
            Coma & $14.84 \pm 0.11$ & $14.91 \pm 0.11$\tnote{1} \\
            A1689 & $14.88 \pm 0.10$ & $15.05 \pm 0.12$\tnote{2} \\
            A85 & $14.78 \pm 0.13$ & $14.88 \pm 0.07$\tnote{3} \\
            A119 & $14.60 \pm 0.12$ & $14.61 \pm 0.11$\tnote{4} \\
            A576 & $14.72 \pm 0.10$ & $14.69 \pm 0.08$\tnote{3} \\
            A1651 & $14.85 \pm 0.13$ & $14.81\pm0.11$\tnote{3} \\
            A2142 & $14.83 \pm 0.14$ & $14.95^{+0.04}_{-0.14}$\tnote{\hspace{0.21cm}5} \\
            A2670 & $14.60 \pm 0.10$ & $14.72 \pm 0.10$\tnote{4} \\
            \hline
            \end{tabular}
            \begin{tablenotes}
                \item[1] \citet{lokas2003coma}
                \item[2] \citet{Lemze2009}
                \item[3] \cite{Wojtak2007_clusters}
                \item[4] \citet{Abdullah2020}
                \item[5] \citet{Munari2014}
            \end{tablenotes}
        \end{threeparttable}
    \end{center}
\end{table}

\begin{figure}
    \centering
    \includegraphics[width=\hsize]{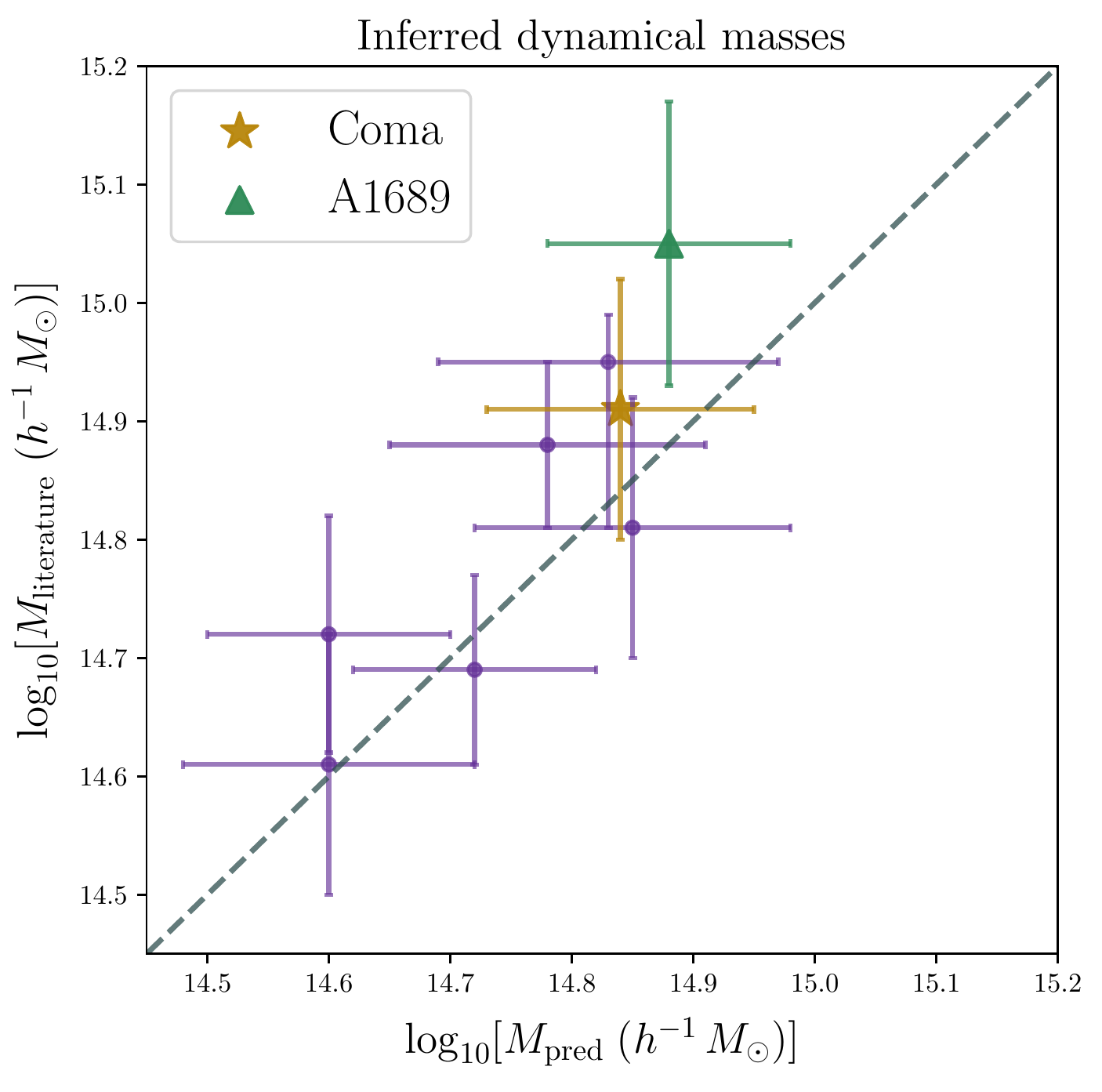}
    \caption{Comparison between mass measurements obtained with our NF mass estimator and the reference mass estimates from the literature, based on the Jeans analysis or the virial mass estimator (cf. Table~\ref{tab:inference_real_clusters}), for eight clusters of galaxies. The error bars represent their corresponding $1\sigma$ uncertainties. The golden star indicates the inferred mass of the Coma cluster.}
    \label{fig:real_cluster_comparison}
\end{figure}

\begin{figure*}
	\centering
		{\includegraphics[width=\hsize,clip=true]{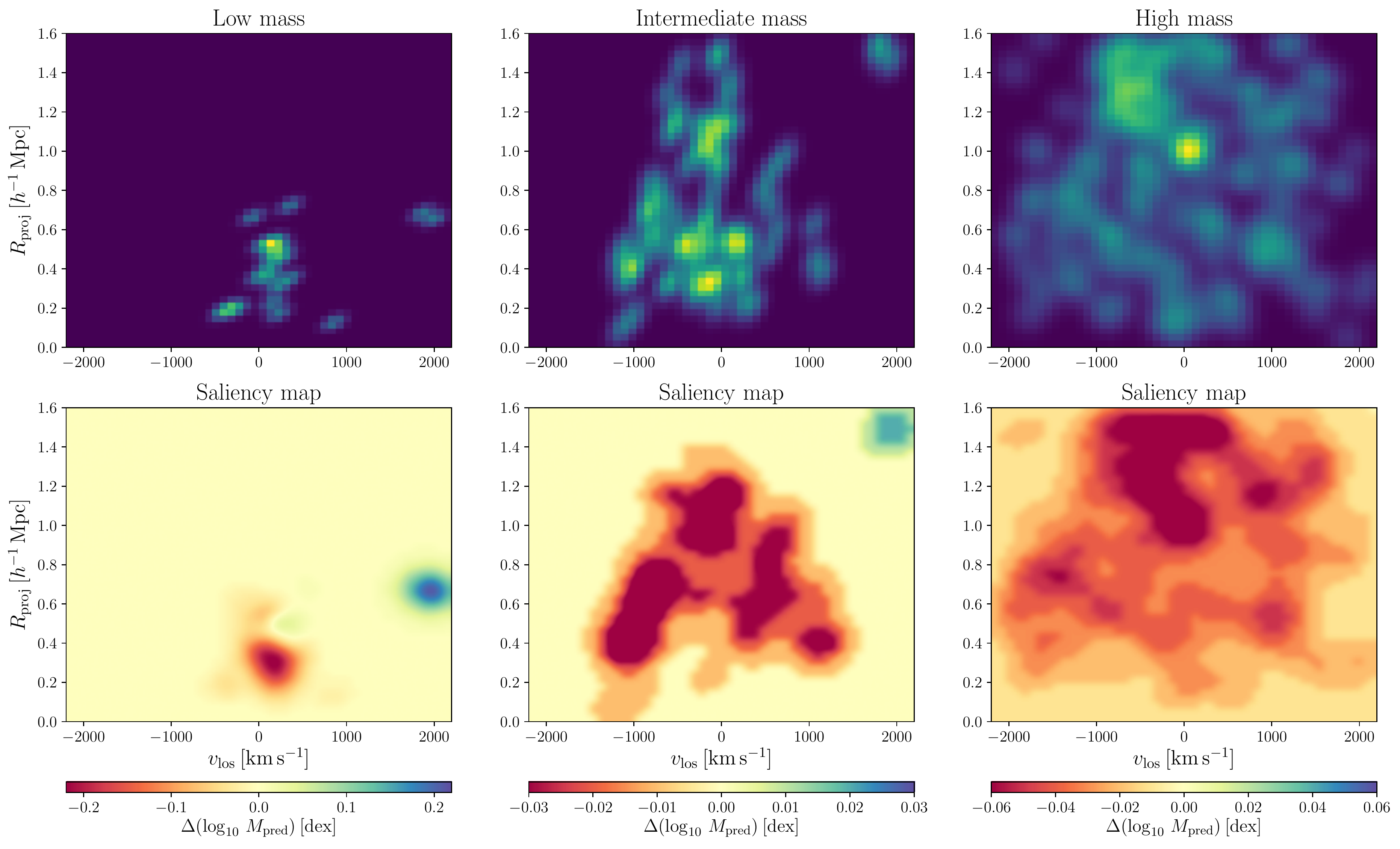}}
	\caption{Saliency maps illustrating the most informative structures in the input 2D phase-space distributions. {\it Top panels:} The 2D Gaussian KDE representation for the phase-space distributions of three clusters in different mass regimes, which are fed as inputs to the NF mass estimator. {\it Bottom panels:} The corresponding saliency maps, which depict the change in network mass prediction when scanning the input image with a Gaussian mask centered at each pixel. The saliency maps are a topographical representation of cluster regions (in red) with significant contributions to the mass estimate. However, the regions (in blue) plausibly related to galaxy interlopers are properly accounted for by the neural network in the mass estimation, showing that the network has learned to inherently account for spurious contributions from interlopers to a reasonable extent.}
	\label{fig:saliency_map}
\end{figure*}

\medskip
Using the NASA/IPAC Extragalactic Database,\footnote{\url{https://ned.ipac.caltech.edu}} we compile all spectroscopic redshifts of galaxies observed in the field of the selected galaxy clusters. The primary sources of redshifts measurements include the 2dF Galaxy Redshift Survey \citep{Colles2001}, the Sloan Digital Sky Survey \cite{SDSS2009}, NOAO Fundamental Plane Survey \citep{Smith2004}, the ESO Nearby Abell Cluster Survey \citep{Katgert1998},  catalogue from \citet{Durret1998} and redshift data for A1689 from \citet{Lemze2009,Czoske2004}. The reference cluster mass measurements include the most updated dynamical estimates based on the Jeans analysis \citep{lokas2003coma,Wojtak2007_clusters,Munari2014} and the virial mass estimator \citep{Abdullah2020}. They were converted to the overdensity parameter adopted in this study.

\medskip
The individual galaxy clusters are preprocessed in a similar way to the preparation of the training set (cf. Section~\ref{dynamical_phase_space_distribution}), i.e. we compute the 2D Gaussian KDE of their respective phase-space distributions. The 2D KDE distributions are then fed to our NF mass estimator, resulting in the inferred dynamical masses listed in Table~\ref{tab:inference_real_clusters}. The masses are indicated as $\log_{10}[M_{\mathrm{200c}} \, (h^{-1} M_{\odot})]$, with the literature values also shown for comparison. The inferred dynamical masses are remarkably in agreement with previous measurements, as illustrated in Fig.~\ref{fig:real_cluster_comparison}, although such network predictions rely on the accuracy and fidelity of the mock observations in the training set to replicate actual observations \citep[e.g.][]{cohn2020multiwavelength}. Nevertheless, the fact that the network can reproduce well-established masses, such as for the Coma cluster, further showcases the efficacy of our NF mass estimator. Since we do not apply any interloper removal scheme to the contaminated catalogue for training nor the above real cluster observations, this implies that the neural flow is capable of building an internal representation of features to account for interlopers to an adequate extent. While a detailed investigation of the cluster masses is beyond the scope of this work, these preliminary results are very promising and demonstrate the potential of neural flows for dynamical mass inference from next-generation galaxy surveys.

\section{Interpreting the model performance}
\label{model_interpretation}

To interpret the performance of our neural flow model, we derive saliency maps for the three clusters in different mass regimes. A saliency map is a topographical representation of the informative features in a given input image. To this end, we study the inferred mass, i.e. the maximum {\it a posteriori} estimate, when specific regions of the input image, i.e. the 2D KDE representation of the phase-space distribution, are masked.

\medskip
Following the approach adopted in \citet{yan2020galaxy}, we apply a Gaussian mask to every pixel in the $50\times50$ image plane to generate 2500 images with distinct excised regions for a particular cluster. The 2D Gaussian mask is defined by
\begin{equation}
    \mathcal{R}_{ij} \equiv 1 - \exp \left[ - \frac{(i - a)^2 + (j - b)^2}{2\sigma^2} \right] ,
    \label{eq:gaussian_mask}
\end{equation}
where $(a, b)$ denotes the centre of the Gaussian mask in image coordinates $(i, j)$. We opt for $\sigma = 2$ pixels, which corresponds to $v_{\mathrm{los}} = 176 \: \mathrm{km} \, \mathrm{s}^{-1}$ and $R_{\mathrm{proj}} = 0.064 \: h^{-1} \, \mathrm{Mpc}$. We subsequently infer the dynamical masses for the set of 2500 images using our trained model and compute the mass difference, $\Delta (\log_{10} M_{\mathrm{pred}} ) = \log_{10} (M_{\mathrm{mask}} / M_{\mathrm{pred}})$, where $M_{\mathrm{mask}}$ refers to the mass for the masked image.

\medskip
We perform this experiment for the three clusters in Fig.~\ref{fig:robustness_tests}, with the resulting $\Delta (\log_{10} M_{\mathrm{pred}} )_{ij}$ maps displayed in the bottom panels of Fig.~\ref{fig:saliency_map}. For comparison, the top panels depict the input 2D KDE images for the different clusters. When masked, the pixels with a substantial contribution to the original mass estimate will yield negative values in this difference map. As expected, we find that the dense regions of the phase-space distribution, which are associated with the main central cluster, contribute most significantly to the dynamical mass prediction. In contrast, distant structures from the cluster centre, as in the case of the clusters with low and intermediate masses, are taken into account by the neural network in the mass estimation. These regions are likely to correspond to the presence of interlopers. This, therefore, shows that the network has learned to account for the presence of interloper galaxies to a reasonable extent, highlighting another appealing aspect of our neural flow approach. With the identification of interlopers being a highly non-trivial task in practice, this justifies the development and application of such deep learning machinery to the cluster mass estimation problem. Moreover, the saliency maps seem to indicate that part of the information on cluster masses potentially emanate from substructures apparent in the phase-space diagrams. In this case, the NF mass estimator would be able to correct a bias related to the presence of dynamical substructures \citep{old2018clusterIII}.

\section{Conclusions and outlook}
\label{conclusions}

We have presented a first attempt at dynamical mass inference of galaxy clusters with robust uncertainty quantification using neural flows. We obtain very promising results by fitting our model to the 2D joint phase-space distribution of galaxy clusters, which consists of the projected radial distance, $R_{\mathrm{proj}}$, from cluster centre and the galaxy line-of-sight velocity, $v_{\mathrm{los}}$. Our neural network architecture is inspired by the novel advances pertaining to sophisticated neural density estimators. We have employed a normalizing flow with relatively few parameters, such that the model can be trained within a few minutes on a standard GPU, with the subsequent predictions of cluster mass PDFs being nearly instantaneous.

\medskip
Our neural network predictions have a mean overall logarithmic residual scatter of 0.028~dex when applied to a test set contaminated with interloper galaxies, with a log-normal scatter of 0.126~dex, which goes down to 0.089~dex for the mass range considered by \citet{ho2019robust}. This constitutes an improvement by around $33 \%$ over their recently developed CNN, which, in turn, is a factor of three improvement over the classical $M-\sigma$ scaling relation. As such, our neural flow mass estimator yields an improvement by nearly a factor of four relative to such scaling relations, while outperforming other recent machine learning approaches.

\medskip
To demonstrate the potential of our neural network, we have applied the trained model to a selection of real galaxy clusters and infer their corresponding dynamical masses. We find that the neural network performs remarkably well, yielding inferred masses consistent with past measurements available from literature (cf. Table~\ref{tab:inference_real_clusters}). This undoubtedly serves to highlight the efficacy of our neural flow model, showcasing it as an extremely promising tool for dynamical mass inference from upcoming galaxy surveys. In an attempt to introspect the model performance, we have derived saliency maps to visualize the most informative regions of the phase-space distribution. We find that the neural network inherently accounts for structures plausibly linked to interlopers to some extent, which is another interesting aspect of our approach. Such sophisticated techniques would undoubtedly be relevant for robust and efficient dynamical mass inference from upcoming surveys covering unprecedented volumes of the sky.

\medskip
While we have implemented a network of relatively low complexity, there are, nevertheless, a series of interesting possibilities with a further level of sophistication that are worth exploring. Instead of viewing the problem in a conditional density estimation setting, we can use neural autoregressive flows to model the likelihood, $\mathcal{P}( \{ \myvec{R}_{\mathrm{proj}}, \myvec{v}_{\mathrm{los}} \} | M) \sim \mathcal{P}( \hat{\myvec{d}} | M)$, where $\hat{\myvec{d}}$ corresponds to a set of informative and sufficient summary statistics extracted by a separate neural network, such as the Information Maximizing Neural Network (IMNN) developed by \citet{charnock2018IMNN}. The use of the IMNN would circumvent the use of standard summary statistics such as the velocity dispersion $\sigma_{\mathrm{v}}$, which, as mentioned previously, does not account for all the relevant physical effects, such that it is inadequate in practice. This would, subsequently, allow us to work in a Bayesian framework to infer the desired posterior as $\mathcal{P}(M | \hat{\myvec{d}}) \propto \mathcal{P}( \hat{\myvec{d}} | M) \times \mathcal{P}(M)$, where $\mathcal{P}(M)$ is the corresponding prior on the dynamical mass $M$ which could be physically motivated to account for effects such as presence of interlopers.

\medskip
Another plausible and exciting avenue is to make use of a conditional pixelwise probability estimator \citep{lanusse2019pixel} to work directly at the level of 3D cluster dynamics, i.e. $\{ \myvec{x}_\mathrm{proj}, \myvec{y}_\mathrm{proj}, \myvec{v}_{\mathrm{los}} \}$. By utilizing 3D convolutions in this uncompressed space, the neural network will, in principle, inherently account for the presence of interlopers in a given galaxy cluster more effectively than the NF mass estimator presented in this work. As we develop more sophisticated neural mass inference algorithms, it also becomes essential to model the (epistemic) uncertainty associated with the weights of the neural network \citep{ho2020approximate} to better account for the regions of the parameter space with scarce training data (such as the most massive galaxy clusters in a catalogue). To this end, it is worth exploring recently developed techniques involving variational inference to model the network weights with some probability distribution \citep[e.g.][]{blundell2015weight, wen2018flipout} instead of using only their maximum likelihood estimates when performing the mass inference. By eventually marginalizing over the learned distributions of the weights, we would ensure that the uncertainties associated to the inferred masses are not underestimated.

\section*{Acknowledgements}

We express our appreciation to the reviewer whose constructive feedback helped to improve various aspects of our method. We convey our gratitude to Matthew Ho for providing us with the mock catalogues. We thank Gary Mamon, Michelle Ntampaka, Guilhem Lavaux and Tom Charnock for their comments on our manuscript. DKR is a DARK fellow supported by a Semper Ardens grant from the Carlsberg Foundation (reference CF15-0384). This work was supported by a VILLUM FONDEN Investigator grant (project number 16599). CG was supported by a VILLUM FONDEN Young Investigator grant (project number 25501). This work has made use of the Horizon and Henon Cluster hosted by Institut d'Astrophysique de Paris. This work has been done within the activities of the Domaine d'Int\'er\^et Majeur (DIM) Astrophysique et Conditions d'Apparition de la Vie (ACAV), and received financial support from R\'egion Ile-de-France. This work is done within the Aquila Consortium.\footnote{\url{https://aquila-consortium.org}}

\section*{Data availability}

The data underlying this article will be shared on reasonable request to the corresponding author.




\bibliographystyle{mnras} 
\bibliography{./compiled_references} 



\appendix

\section{Gaussian kernel density estimation}
\label{appendix_KDE}

A comprehensive review of kernel density estimation is provided in \citet{diggle1984monte, wand1994kernel, sheather2004density}. For a set of $n$ independent and identically distributed univariate samples $\{ \myvec{X}_1, \myvec{X}_2, \ldots, \myvec{X}_n \}$, where each variable $\myvec{X}_i$ is described by ($v_{\rm{los}}, \, R_{\rm{proj}}$), drawn from an unknown distribution, the density $f$ evaluated at a given point $\myvec{x} = (v_{\rm{los}}, \, R_{\rm{proj}})$ can be estimated using a kernel density estimator via
\begin{equation}
\hat{f}(\myvec{x}) = \frac{1}{n | \mathbf{H} |^{1/2} } \sum^n_{i=1} \, K \big[ \mathbf{H}^{-1/2} (\myvec{x} - \myvec{X}_i) \big],
\end{equation}
where $K$ is a non-negative kernel function and $\mathbf{H}$ is a $2\times2$ bandwidth matrix. The kernel density estimator performs a sum of the density contributions from the series of data points $\{ \myvec{X}_1, \myvec{X}_2, \ldots, \myvec{X}_n \}$ at the evaluation point $\myvec{x}$. The choice of kernel function determines the shape of the density contributions, with their size and orientation characterized by the bandwidth matrix. In this work, we use a 2-dimensional (or bivariate) Gaussian kernel function described by
\begin{equation}
K(\myvec{u}) = (2 \pi)^{-3/2} \, |\mathbf{H}|^{-1/2} \; \exp \left(- \tfrac{1}{2} \, \myvec{u}^\intercal \, \mathbf{H}^{-1} \, \myvec{u} \right),
\end{equation}
where $\myvec{u} = \myvec{x} - \myvec{X}_i$. For the bandwidth matrix, a constant coefficient is multiplied with the covariance matrix of the data, $\mathbf{H} = h_0 \mathbf{\Sigma}$, where the bandwidth scaling factor is taken to be $h_0 = 0.15$, which ensures the robustness of our method to cluster richness and typical velocity errors involved in spectroscopic observations, as justified in Section~\ref{validation_performance}.

\bsp	
\label{lastpage}
\end{document}